\documentclass[12pt]{article}

\usepackage[perpage]{footmisc}
\usepackage[T1]{fontenc}
\usepackage{amsmath}
\usepackage{amssymb}
\usepackage{authblk}
\usepackage{slashed}
\usepackage{amsfonts}
\usepackage{cite}
\usepackage{graphicx}
\usepackage[dvipsnames]{xcolor}
\usepackage[colorlinks=true,urlcolor=blue,linkcolor=blue,citecolor=blue]{hyperref}
\usepackage{cancel}
\usepackage{booktabs}
\usepackage{caption}
\usepackage{subcaption}
\usepackage{float}

\usepackage[top=1.135in,bottom=1.17in,left=1.16in,right=1.16in]{geometry}
\numberwithin{equation}{section}

\allowdisplaybreaks

\newcommand{\be}{\begin{equation}}
\newcommand{\ee}{\end{equation}}

\begin{document}
\thispagestyle{empty}

\title{Dirac-Majorana neutrinos distinction in four-body decays.}

\vspace{50pt}

\author[1]{Juan Manuel Márquez}
\author[1]{Diego Portillo-Sánchez}
\author[1]{Pablo Roig}

\affil[1]{Departamento de F\'isica, Centro de Investigaci\'on y de Estudios Avanzados del Instituto Polit\'ecnico Nacional\\
Apartado Postal 14-740, 07360 Ciudad de M\'exico, M\'exico}

\maketitle
\abstract{
A novel method to differentiate the effects of Dirac and Majorana (D-M) neutrinos in four-body decays has been discussed in Ref.~\cite{Kim}. There, it is concluded that the back-to-back kinematic scenario seems to avoid the constraint imposed by the ‘practical Dirac-Majorana confusion theorem’, as one does not need to fully integrate over neutrino and antineutrino momenta. In this paper, we propose to analyse radiative leptonic lepton-decays ($\ell\to\ell'\nu\bar{\nu}\gamma$), as an independent alternative process to study the possible Majorana nature of neutrinos.
Our approach demonstrates that, in the back-to-back kinematic configuration (for the $\ell'- \gamma$ and $\nu-\bar{\nu}$ systems, respectively), the distinction between Dirac and Majorana cases disappears when the inaccessible neutrino angle is integrated out, which might appear unexpected considering the claims in \cite{Kim}. Our results apply in absence of non-standard interactions, which can enhance generally the sensitivity to the neutrino nature.}

\section{Introduction}
\label{sec:intro}
Nowadays the Dirac or Majorana nature of neutrinos, as well as the mechanism from which they acquire mass, is one of the unsolved puzzles whose solution seems to lie beyond the Standard Model (SM). There are many minimal extensions of it in order to account for nonzero masses and mixings for the active neutrinos; namely adding new gauge singlet fields, such as right-handed neutrinos. Some of them are the well-known $\nu SM$ \cite{PMNS1,PMNS2} and the seesaw-mechanisms \cite{SS1,SS2,SS3,SS4,SS5}.

If neutrinos were Majorana particles, lepton number would  not be conserved and neutrinos would be their own antiparticles, i.e,  no conserved quantum number would distinguish  neutrino and anti-neutrino. There are some important proposals to probe the Majorana nature of the neutrino, such as the neutrino-less double beta decay $(0\nu\beta\beta)$ 
of nuclei
~\footnote{This could however be unobservable, even if neutrinos are Majorana particles, depending on their properties, like the neutrino mass ordering.} \cite{Rodejohann:2011mu,Vergados:2012xy,Barea:2013bz,Engel:2016xgb,Dolinski:2019nrj}, coherent scattering of neutrino on nucleus with bremsstrahlung radiation \cite{Millar:2018}, and many other lepton number violating processes \cite{LFV,Barbero:2002wm,Barbero:2007zm,Barbero:2013fc,Ilakovac:1995wc, Gribanov:2001vv, Helo:2010cw, LopezCastro:2012udb, Yuan:2017xdp, Kim:2017pra, LopezCastro:2012rbs,Cvetic:2010rw,Quintero:2011yh, Yuan:2013yba, Castro:2013jsn, Cvetic:2014nla, Cvetic:2015ura, Cvetic:2015naa, Mandal:2016hpr, Moreno:2016cfz, Cvetic:2016fbv, Cvetic:2017vwl, Cvetic:2017gkt, Yuan:2017uyq, Li:2018pag, Wang:2018bgp, Kim:2019xqj, Cvetic:2020lyh,Mejia-Guisao:2017nzx, Das:2021prm, Das:2021kzi, Zhang:2021wjj,Hernandez-Tome:2021byt,Hernandez-Tome:2022ejd},  most of them motivated by the propagation of a massive heavy neutral lepton.

However, the difference among  
Dirac and Majorana neutrinos, when the unobservable neutrinos momenta get fully integrated out, is proportional to some power of neutrino masses, assuming they enter left-handed charged weak currents, as in the SM. This is precisely  stated by the “practical Dirac-Majorana confusion theorem” (DMCT) \cite{Kayser}, and makes extremely challenging to distinguish experimentally between both possibilities, specially for tiny neutrino masses.

It is crucial that the difference between Dirac and Majorana nature can be settled independently of the mass of the neutrinos provided their momenta are not integrated out. Since neutrinos momenta are not experimentally accessible, we need a method to infer  them, so that we can discern the neutrinos variables without the need of any explicit observation of them, which is extremely difficult any way at present.  Kim, Murthy and Sahoo claimed \cite{Kim} that we can deduce the neutrinos momenta working in the back-to-back ($b2b$ from now on) kinematic configuration of a four-body decay with two final-state neutrinos, where the difference between Dirac and Majorana cases does survive irrespectively of the neutrino mass values, as long as neutrinos are not strictly massless.

This could be a really exciting and important strategy to distinguish the specific neutrino nature. Motivated by this method (“KMS method” from now on), we analyse radiative leptonic lepton-decays ($\ell^-\rightarrow \ell^{\prime}{}^-\nu_{\ell}
\bar{\nu}_{\ell^{\prime}}
\gamma$) as an independent approach in order to distinguish the Dirac or Majorana nature of neutrinos. We emphasize that the quoted KMS method of analysis considers  processes with $\nu\bar{\nu}$ final states with same-flavour neutrinos, in order to apply the quantum statistical properties of Majorana neutrinos. Nevertheless, the Majorana nature of neutrinos could be implemented even with $\nu\bar{\nu}$ final states with different flavours, working in the mass basis, where the quantum statistical properties could be applied to distinguish between Dirac and Majorana nature, as we shall discuss, see e.g.  \cite{Kotani,Shrock1980,Shrock1982,MichelParameters}~\footnote{Obviously, physics conclusions cannot depend on basis choice, although a particular one can be most convenient in a given analysis.}. This fact allows us to study many other processes as an alternative strategy (without hadronic transitions) with larger branching ratios (BR), such as the radiative leptonic decays of leptons, in which we focus here, that we will study neglecting possible non-standard interactions. 

In Ref. \cite{Rosen}, it was suggested that measurements of the Michel parameters in muon decay cannot help to distinguish the nature of neutrinos because they are of different flavours and the required anti-symmetrization of the amplitude for Majorana neutrinos does not apply. However, as we shall discuss, the reason why we cannot distinguish between Dirac and Majorana nature of neutrinos in this process was carefully explained  by London and Langacker in \cite{London}; they have shown that it is not possible, even in principle, to test lepton-number conservation in muon decay if the final neutrinos are massless and are not observed (which agrees with the DMCT theorem for the SM neutrinos interaction).  
Therefore, if we can infer some information on neutrinos, it becomes possible to get a non-vanishing difference between Dirac and Majorana cases, even for this type of leptonic decays. 

As it was first introduced in \cite{Shrock1980}, and discussed in several following works \cite{Shrock1981,Kotani,Shrock1982,MichelParameters}, the Majorana nature could affect processes with $\nu \Bar{\nu}$ final states involving different flavours. We emphasize that this effect is not a direct consequence of the presence of indistinguishable fermions in the final state but a result of the intrinsic Majorana fermion properties. Using the fact that flavoured neutrinos do not have a definite mass; i.e., the existence of two different basis, due to finite neutrino masses, this could lead to “internal interference effects” \cite{London} when considering Majorana neutrinos. These effects can be analyzed with the usual implementation of Feynman diagrams and their corresponding Feynman rules for Majorana fermions. These interference effects can be explained in detail from first principles by studying the action of Majorana fields over the final asymptotic neutrino states in the QFT scheme, where now the operator $\Psi$ can create either of the two states, and so can $\Bar{\Psi}$.

The main idea is now clear: we will study the possible effects of Majorana neutrinos in the radiative leptonic decay of $\mu$ and $\tau$ leptons assuming we can `measure' them (i.e. infer their kinematic variables), avoiding  integration over their momenta. In such cases, we shall prove that the difference between Dirac and Majorana nature of neutrinos is still present at the level of differential decay rate and explicitly depends on the neutrinos kinematics. Then, given the $b2b$ configuration analysis as explained in Section \ref{subsec:3.2}, we will show that the D-M difference vanishes upon integration on the neutrinos variables and discuss the origin of the discrepancy with the KMS result.

This article is structured as follows: in section \ref{sec:2} we present the explicit radiative lepton-decay rate in the $b2b$ configuration, taking into account Dirac (section \ref{subsec:2.1}) and Majorana (section \ref{subsec:2.2}) neutrinos. In section \ref{subsec:3.1} we present the final energy and angular spectrum obtained from applying our approach. The main results are discussed: specifically showing that the differences for distinct neutrino nature do not survive independently of the non-vanishing neutrino mass, once the unobservable neutrino angle is integrated out. Then, in section \ref{subsec:3.2}  we track the origin of the absence of a difference between Dirac and Majorana cases in the radiative leptonic lepton decay, we elaborate on 
and clarify it, with the implementation of consistency tests and helpful discussions.
Finally our conclusions are given in section \ref{sec:4}. Very useful complementary computations of the phase space treatment, the $b2b$ configuration and the specific branching ratio calculation can be found in Appendix \ref{app:phasespace}, \ref{app:b2b} and \ref{app:br}, respectively,  where all the comments raised on Ref.\cite{Kim2} are corrected and clarified in detail.

\section{Radiative leptonic $\ell$-decay}
\label{sec:2}

\begin{figure}[tbp]
\begin{tabular}{cc}
 \includegraphics[scale=.45]{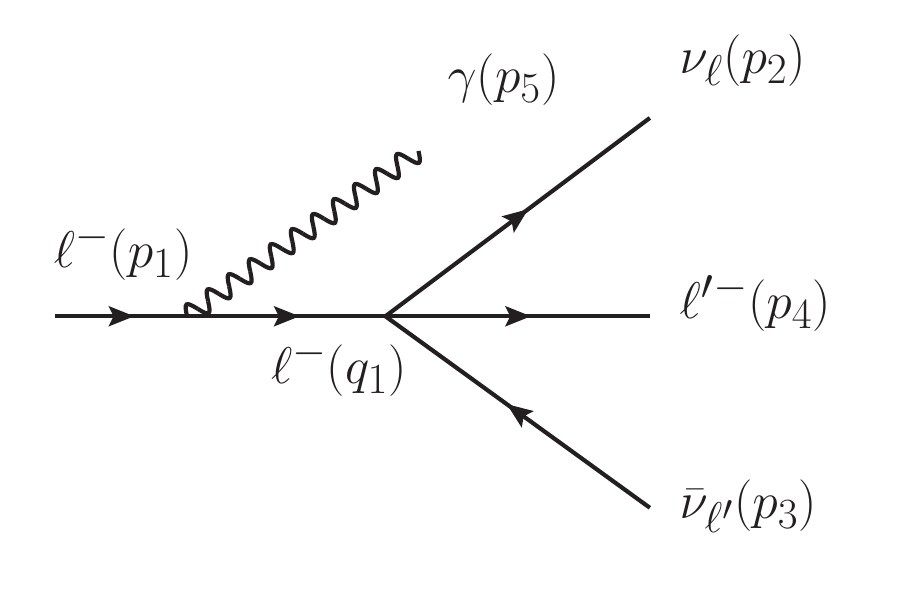} & \includegraphics[scale=.45]{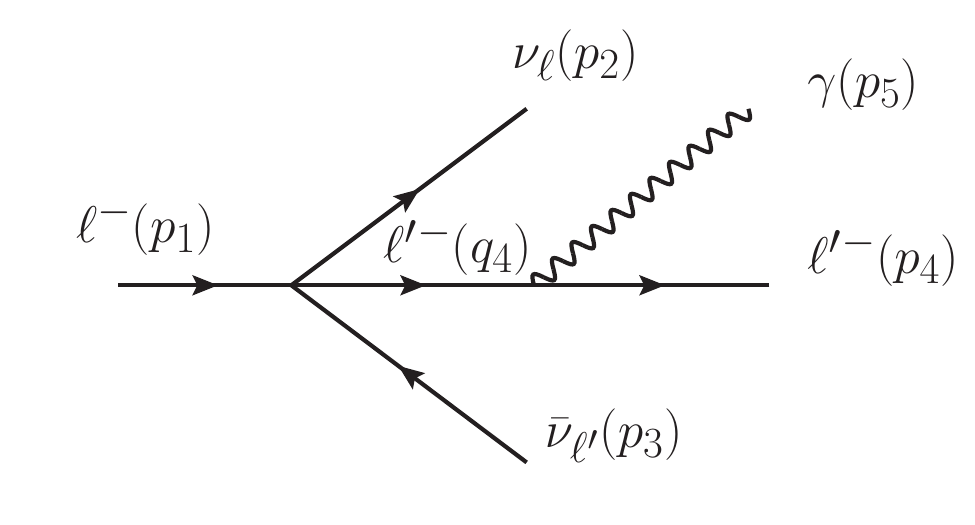} \\
 {\small (a)} & {\small(b)} \\
 \end{tabular}
 \caption{Lowest order Feynman diagrams for radiative leptonic lepton-decay in the Dirac neutrino case.}
 \label{fig:diagrams}
 \end{figure}

Since $m_\ell<<M_W$, we can safely integrate out the $W$ boson and use the Fermi-type  theory of weak interactions to describe the charged lepton decays.
Then, the leading Feynman diagrams contributing to the radiative leptonic $\ell$-decay are shown in figure \ref{fig:diagrams}. 
 
From now on, we will be working in the basis where the mass matrix of charged leptons is already diagonalized and the flavor-eigenstate neutrino ($\nu_{L}$) is taken to be the superposition of the mass-eigenstate neutrinos ($N_j$) with mass $m_j$, that is,
\begin{equation}
    \nu_{\ell L}=\sum_j U_{\ell j}N_{jL},
    \label{changebase}
\end{equation}
where $j=\{1,2,3,..,n\}$ is tagging   mass-eigenstate neutrinos.

In the mass basis, the $\ell^-\rightarrow \ell^{\prime}{}^-\nu_{\ell}
\bar{\nu}_{\ell^{\prime}}
\gamma$ decay consists of the subsets of all the $n^2$ separate decays of neutrino mass eigenstates allowed by phase space, i.e., the incoherent sum of the separate modes $\ell^{-}\to \ell'^{-}\overline{N}_j N_k \gamma$ \cite{Shrock1980}:
\begin{equation}
    d\Gamma(\ell^-\rightarrow \ell^{\prime}{}^-\nu_{\ell}
\bar{\nu}_{\ell^{\prime}}
\gamma)=\sum_{j,k}d\Gamma(\ell^{-}\to \ell'^{-}\overline{N}_j N_k \gamma)
\end{equation}
Note that $\overline{N}$ represents an antineutrino for the Dirac neutrino case, but should be identified with $N$ for the Majorana neutrino case ($N$=$N^c$=$C\overline{N}^T$).

Here it is important to remark that all the following analysis and results are valid only for massive neutrinos, no matter how light their masses could be, as long as they are nonzero. In this case, we have two different well defined basis (mass and flavor) and the corresponding neutrino nature will affect in a different way the differential decay rate computation as we shall see. In the case of massless neutrinos, only one basis exists and then eq.(\ref{changebase}) does not hold, also the $U(1)_{B-L}$ symmetry of the Lagrangian is recovered and our results do not apply. For a helpful discussion see ergo \cite{Pal:2010ih}.

\subsection{Dirac Neutrino Case}
\label{subsec:2.1}
In the Dirac case, the corresponding amplitude for the process $\ell^{-}\to \ell'^{-}\overline{N}_j N_k \gamma$ is given by:\footnote{Where, when computing the decay rate, we must sum  incoherently over the probabilities of all the allowed \{$j$, $k$\} channels.}
 \begin{equation}
     \mathcal{M}^D=\mathcal{M}_{(a)}+\mathcal{M}_{(b)}\equiv\mathcal{M}(p_2,p_3),
 \end{equation}
 where (see figure \ref{fig:diagrams} for momenta convention)
 \begin{equation}
 \begin{split}
     &\mathcal{M}_{(a)}=U_{\ell' j}U_{\ell k}^{*}\frac{eG_F }{\sqrt{2}}\left[\bar{u}_4\gamma^\mu(1-\gamma^5)v_3 \right]\bar{u}_2\gamma_\mu(1-\gamma^5)\left(\frac{\slashed q_1 +m_1}{q_1^2-m_1^2} \right)\gamma^\nu \epsilon^{*}_\nu u_1,\\
     & \mathcal{M}_{(b)}=U_{\ell' j}U_{\ell k}^{*}\frac{eG_F }{\sqrt{2}}\bar{u_4}\gamma^\nu \epsilon^{*}_\nu \left(\frac{\slashed q_4 +m_4}{q_4^2-m_4^2} \right)\gamma_\mu(1-\gamma^5) v_3\left[\bar{u}_2\gamma^\mu(1-\gamma^5)u_1 \right].
 \end{split} 
 \end{equation}
 
 Neglecting all  final lepton masses, as a good approximation, the unpolarized squared amplitude is:
 \begin{equation}
     \begin{split}
        |\overline{\mathcal{M}^D}|^2&=|U_{\ell' j}U_{\ell k}^{*}|^2\frac{64 e^2 G_F^2}{(p_4\cdot p_5)(p_1\cdot p_5)^2} \bigg\{ (p_1\cdot p_3) \Big[(p_2\cdot p_4+p_2\cdot p_5)(p_1\cdot p_5)^2-m_1^2(p_2\cdot p_4)(p_4\cdot p_5)\\
        &+(p_1\cdot p_4)(p_1\cdot p_5)(2(p_2\cdot p_4)+p_2\cdot p_5)-(p_1\cdot p_5)(p_4\cdot p_5)(p_1\cdot p_2+p_2\cdot p_4)\Big]+(p_2\cdot p_4)\\
        &\Big[(p_3\cdot p_4)(p_1\cdot p_5)^2+m_1^2 (p_3\cdot p_5)(p_4\cdot p_5)+(p_1\cdot p_5)(p_3\cdot p_5)(p_4\cdot p_5 - p_1\cdot p_4)\Big]\bigg\}.
     \end{split}
 \end{equation}
 \begin{figure}[tbp]
     \centering
     \includegraphics[scale=.8]{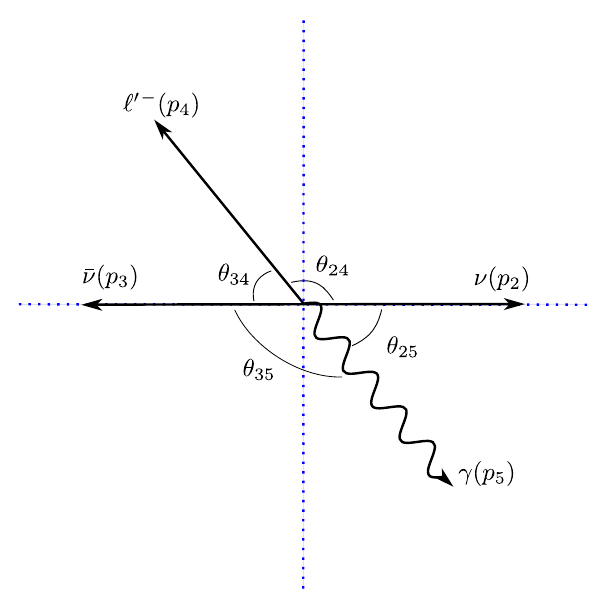}
     \caption{$b2b$ kinematic configuration in the initial charged-lepton rest frame.}
     \label{diag4}
 \end{figure}
 Finally, motivated by the KMS method, we need to work in the initial charged-lepton rest frame, specifically in the $b2b$ kinematic configuration shown in figure \ref{diag4}, where the scalar products, neglecting the final lepton masses, are given by:\footnote{The energies are related by: $E_\gamma=E_{\ell'}$ and $E_\nu=E_{\bar{\nu}} =\frac{m_\ell}{2}-E_{\ell'}$.} 
 \begin{equation}
     \begin{split}
        & p_4 \cdot p_5=2E_{\ell'}^2,\quad  p_2 \cdot p_3=2\left(\frac{m_\ell}{2}-E_{\ell'}\right)^2, \\
         & p_1 \cdot p_2=p_1 \cdot p_3=m_\ell E_{\ell'}\left(\frac{m_\ell}{2}-E_{\ell'}\right), \\
         & p_1 \cdot p_4=p_1 \cdot p_5=m_\ell E_{\ell'}, \\
         & p_3 \cdot p_4=p_2 \cdot p_5=E_{\ell'}\left(\frac{m_\ell}{2}-E_{\ell'}\right)(1+\cos{\theta}), \\
         & p_3 \cdot p_5=p_2 \cdot p_4=E_{\ell'}\left(\frac{m_\ell}{2}-E_{\ell'}\right)(1-\cos{\theta}).
     \end{split}
 \end{equation}
 Thus, in this kinematical configuration, the process is described in terms of only two variables, the final charged-lepton energy\footnote{The corresponding energy range is $0\leq E_{\ell'}\leq\frac{m_\ell}{2}$.}, $E_4\equiv E_{\ell'}$,  and the angle between the neutrino and final charged-lepton directions, $\theta_{24}=\theta$. With these considerations, the final Dirac amplitude is given by:
 \begin{equation}\label{eq:Dbtb}
 |\overline{\mathcal{M}^D_\leftrightarrow}|^2=|U_{\ell' j}U_{\ell k}^{*}|^2\frac{8 e^2 G_F^2(m_\ell-2E_{\ell'})^2}{m_\ell E_{\ell'}}\left(8E_{\ell'}^2\sin^4{\frac{\theta}{2}}+(1+\cos{\theta})m_\ell^2 \right)\, .
 \end{equation}
 The subindex `$\leftrightarrow$' denotes the $b2b$ configuration. The $1/E_{\ell'}$-dependence reflects the infrarred behaviour of the amplitude in the soft-photon limit.
 
\subsection{Majorana Neutrino Case}
\label{subsec:2.2}
 Unlike the Dirac case, the properties of Majorana neutrinos have  strong consequences in the amplitude. For Majorana neutrinos the decay modes $\ell^{-}\to \ell'^{-} \overline{N}_j N_k \gamma$ and $\ell^{-}\to \ell'^{-} \overline{N}_k N_j \gamma$ yield the same final states for $k\neq j$ as well as for $k=j$ (since $\overline{N}_i=N_i$), and hence their amplitudes must be added coherently. This is a result that can be obtained rigorously in the QFT formalism. 
 
\begin{figure}[tbp]
\begin{tabular}{cc}
 \includegraphics[scale=.45]{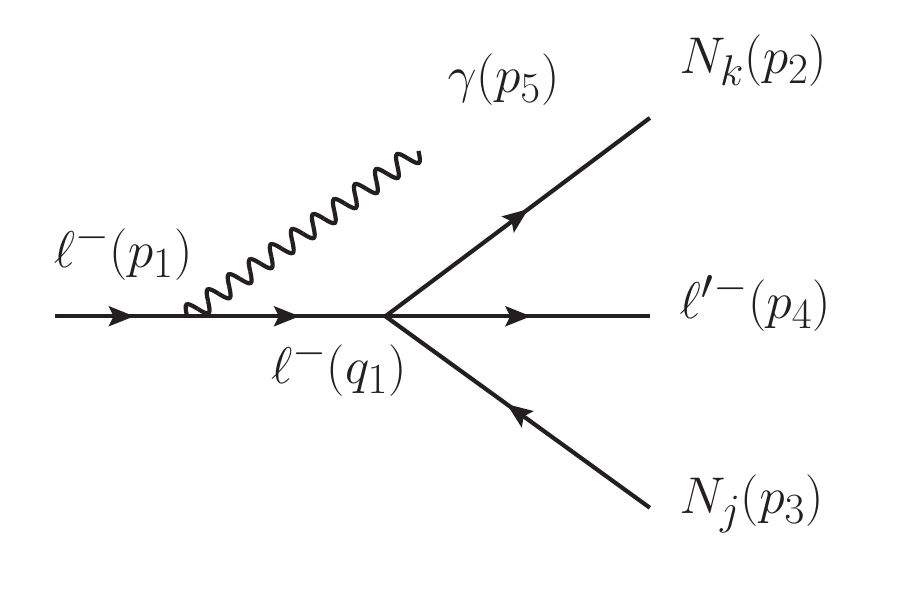} &  \includegraphics[scale=.45]{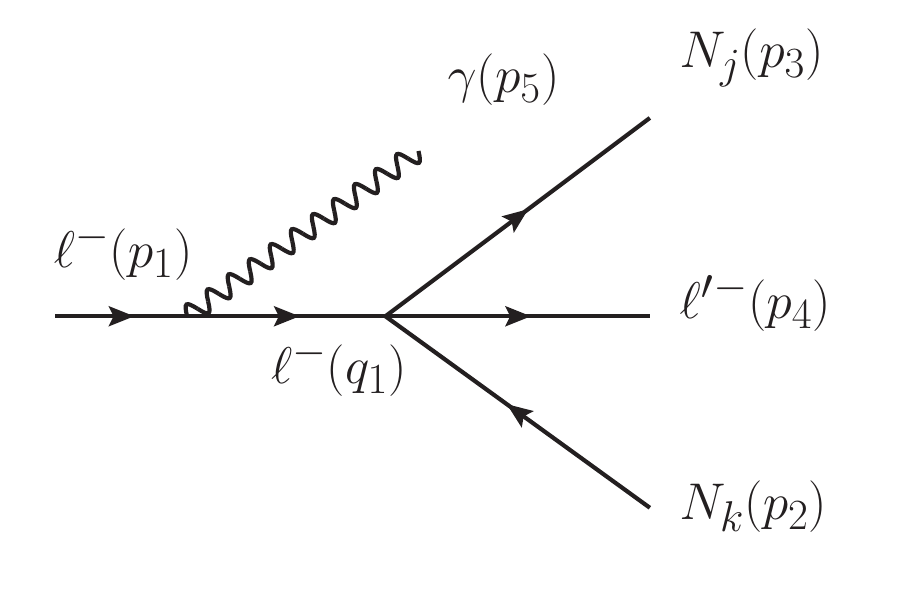} \\
 {\small (a)} & {\small(b)} \\
  \includegraphics[scale=.45]{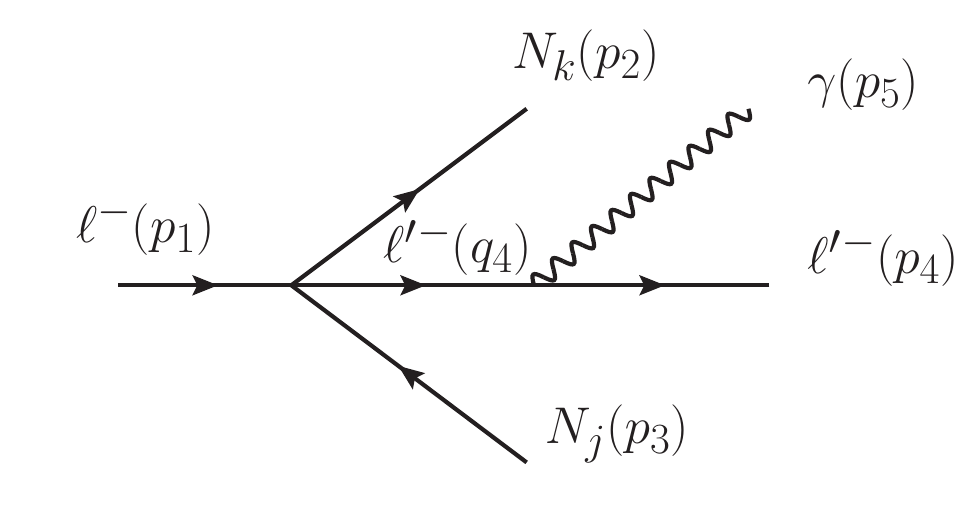} & \includegraphics[scale=.45]{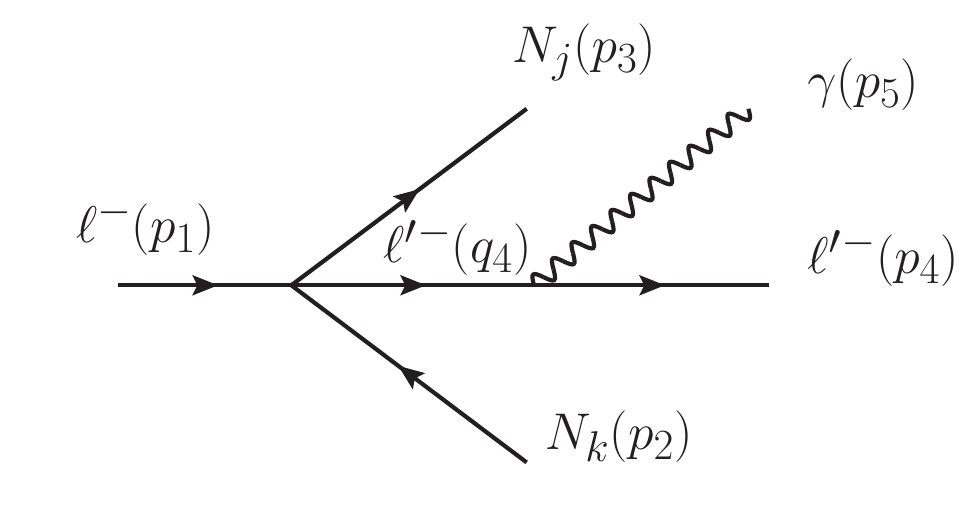} \\
 {\small (c)} & {\small(d)} \\
 \end{tabular}
\begin{tabular}{cc}

 \end{tabular}
 \caption{Lowest order Feynman diagrams for radiative leptonic lepton-decay in the Majorana neutrino case.}
 \label{fig:Majdiagrams}
 \end{figure}

 The possible first order Feynman diagrams for the $\ell^{-}\to \ell'^{-} N_j N_k \gamma$ decay are shown in figure \ref{fig:Majdiagrams}, 
leading to the total amplitude:
 \begin{equation}
 \mathcal{M}^M=\mathcal{M}_{jk}(p_2,p_3)-\mathcal{M}_{kj}(p_3,p_2)\,,
 \end{equation}
 where the relative sign between $\mathcal{M}(p_2,p_3)$ and $\mathcal{M}(p_3,p_2)$ stems from the application of Wick’s theorem in presence of Majorana fermions (see, e.g. Ref.~\cite{Denner}).
 
 It can be shown that, after summing over polarizations, $Re\left(\mathcal{M}(p_2,p_3)\mathcal{M}^{*}(p_3,p_2)\right)\propto m_\nu^2\approx0$ due to the smallness of neutrino masses\footnote{We are considering only the contribution of the three active light neutrinos mass eigenstates.}. Thus
  \begin{equation}\label{eq:CommentAbove}
 |\overline{\mathcal{M}^M}|^2=|\overline{\mathcal{M}}_{jk}(p_2,p_3)|^2+|\overline{\mathcal{M}}_{kj}(p_3,p_2)|^2.
 \end{equation}
 The computation, neglecting the final lepton masses, leads to the following squared amplitude for the $b2b$ kinematic configuration: 
  \begin{equation}\label{eq:Mbtb}
  \begin{split}
 |\overline{\mathcal{M}^M_\leftrightarrow}|^2=\frac{8 e^2 G_F^2(m_\ell-2E_{\ell'})^2}{m_\ell E_{\ell'}}&\left\{|U_{\ell' j}U_{\ell k}^{*}|^2\left(8E_{\ell'}^2 \sin^4\left(\frac{\theta
 }{2}\right)+m_\ell^2(1+\cos \theta)\right)\right.\\
 &\left.+|U_{\ell' k}U_{\ell j}^{*}|^2\left(8E_{\ell'}^2 \cos^4\left(\frac{\theta
 }{2}\right)+m_\ell^2(1-\cos \theta)\right)\right\}\,.
 \end{split}
 \end{equation}
 Thus, the Majorana nature of neutrinos would generate a different behavior of the amplitude compared with the Dirac neutrinos case, eq.~(\ref{eq:Dbtb}). 
\section{Energy and Angular Distributions}
\label{sec:3}
In this section we develop our own derivation, motivated by the KMS method, and compute the corresponding energy and angular distributions of the $\nu-\bar{\nu}$ and $\ell^{\prime}-\gamma$ systems in the $b2b$ configuration, respectively. We will finally discuss our results.

\subsection{Our Approach}
\label{subsec:3.1}

When we restrict ourselves to the $b2b$ scenario, which is just a specific kinematic configuration of the general case, as we did before, we will adopt the following notation of the corresponding decay rate according to eq.~(\ref{diffwidth}) (see Appendix \ref{app:phasespace} for all the details):
\begin{equation}
   \left. \frac{d\Gamma^{D,M}}{dE_{\nu}dE_{\Bar{\nu}}d\cos{\Theta_{\nu\bar{\nu}}}d\cos{\theta_{\ell^{\prime}}}d\phi}\right|_{b2b}= \frac{2 E_{\nu}^2}{ (4\pi)^6 m_{\ell}}\,\frac{1}{\epsilon}\sum_{j,k}|\overline{\mathcal{M}^{D,M}_{\leftrightarrow}}|^2,
   \label{neutrinospectrum}
\end{equation}
\begin{equation}
   \left. \frac{d\Gamma^{D,M}}{dE_{\ell'}dE_{\gamma}d\cos{\Theta_{\ell^{\prime}\gamma}}d\cos{\theta_{\nu}}d\phi}\right|_{b2b}= \frac{2 E_{\ell'}^2}{ (4\pi)^6m_{\ell}}\,\frac{1}{\epsilon}\sum_{j,k}|\overline{\mathcal{M}^{D,M}_{\leftrightarrow}}|^2,
   \label{ellgamaspectrum}
\end{equation}
where $\epsilon=1 (2)$ for Dirac (Majorana) neutrinos, showing that it is just the standard differential decay rate evaluated in the specific $b2b$ kinematics\footnote{Where $E_{\gamma}=E_{\ell^{\prime}}$, $E_{\nu}=E_{\bar{\nu}}$ and $\Theta_{\nu\bar{\nu}}=\Theta_{\ell^{\prime}\gamma}=\pi$.}, and the amplitudes involved were already quoted in the last section.\\
Using the $|\overline{\mathcal{M}^{D/M}_{\leftrightarrow}}|^2$ previously  computed, we obtain the following neutrinos differential decay rate:
   {\small\begin{equation}
  \label{eq:CommentBelow}
   \begin{split}
   &\left.d\Gamma^{D}_{\nu\nu}\right|_{b2b}\equiv\left. \frac{d\Gamma^{D}}{dE_{\nu}dE_{\Bar{\nu}}d\cos{\Theta_{\nu\bar{\nu}}}d\cos{\theta_{\ell^{\prime}}}d\phi}\right|_{b2b}=\frac{ 4 \alpha G_F^2(m_\ell-2E_{\ell^{\prime}})^4}{(4\pi)^5 m_{\ell}^2 E_{\ell^{\prime}}}\left(8E_{\ell^{\prime}}^2\sin^4{\frac{\theta}{2}}+(1+\cos{\theta})m_{\ell}^2 \right),\\
 &\left.d\Gamma^{M}_{\nu\nu}\right|_{b2b}\equiv\left. \frac{d\Gamma^{M}}{dE_{\nu}dE_{\Bar{\nu}}d\cos{\Theta_{\nu\bar{\nu}}}d\cos{\theta_{\ell^{\prime}}}d\phi}\right|_{b2b}=\frac{ 4 \alpha G_F^2(m_{\ell}-2E_{\ell^{\prime}})^4}{ (4\pi)^5 m_{\ell}^2 E_{\ell^{\prime}}}\left(E_{\ell^{\prime}}^2 (3+\cos{2\theta})+m_{\ell}^2 \right),
 \end{split}
 \end{equation}}
 where we already used the unitarity relations for the mixing matrix elements considering just the three active light neutrinos mass eigenstates. We can also consider the presence of a massive sector leaving the sum explicit, where the new sterile neutrinos could be produced if they are energetically allowed, or else could affect the unitarity relation if they are forbidden by kinematics. Nevertheless, as discussed in Ref.~\cite{MichelParameters}, the possible effect would be suppressed by the specific heavy mixing and can not be higher than $10^{-4}$, thus it should only be considered if that precision is needed.

 Then the difference between Dirac and Majorana cases in the $b2b$ scenario is: 
 \begin{equation}
 \label{eqn:difdm}
   \begin{split}
      d\Gamma_{\nu\nu}^D|_{b2b}-d\Gamma_{\nu\nu}^M|_{b2b}&=\frac{4 \alpha G_F^2(m_{\ell}-2E_{\ell^{\prime}})^5}{ (4\pi)^5 m_{\ell}^2 E_{\ell^{\prime}}} \left(m_{\ell} +2 E_{\ell^{\prime}}\right)\cos{\theta}.
  \end{split}
 \end{equation}
The difference is evident, unfortunately, the angle $\theta$ is not experimentally accessible, so we should rather integrate over it to get the final charged-lepton energy distributions for Dirac and Majorana cases.\\
In order to integrate over this inaccessible angle, we just need to rewrite the angle $\theta$ that appears in the right-hand side of the above equations in terms of our phase space variables $\theta_{\ell'}$ and $\phi$, i.e., $\theta=\theta(\theta_{\ell'},\phi)$. \\
Since $\theta$ is the angle between $\vec{p}_{\nu}$ and $\vec{p}_{\ell^{\prime}}$, it is easy to get, as shown in eq.(\ref{eqn:angulargood}) that:
\begin{equation}
    \begin{split}
       \cos{\theta}&\equiv\hat{p}_{4}\cdot\hat{p}_2=\sin{\theta_{\ell^{\prime}}}\sin{\theta_{\nu}}\cos{\phi}-\cos{\theta_{\nu}}\cos{\theta_{\ell^{\prime}}}.
    \end{split}
    \label{eqn:angulargood}
\end{equation}
Now, since this neutrinos differential decay rate does not depend explicitly on $\theta_{\nu}$, we have the freedom to fix it to ease further computations~\footnote{The following discussion can be directly applied to eq.(\ref{ellgamaspectrum}) for $\theta_{\ell^{\prime}}$.} . Then, for example, we can choose the system in such a way that $\theta_{\nu}=0$ and thus we get $\cos{\theta}=-\cos{\theta_{\ell'}}$, which does not have a dependence on the $\phi$ angle, showing explicitly the azimuthal symmetry of this specific selection. Using this dependence, where $0\leq\theta_{\ell'}\leq\pi$ and $0\leq\phi\leq 2\pi$  so that all possible angular configurations between $\ell'$ and $\nu$ are accounted for, we have from eq.(\ref{eqn:difdm}): 
 \begin{equation}
   \begin{split}
  &  \int_0^{2\pi} \int_0^{\pi}\left( d\Gamma_{\nu\nu}^D|_{b2b}-d\Gamma_{\nu\nu}^M|_{b2b}\right)d\theta_{\ell^{\prime}} d\phi\\
    &=\int_0^{2\pi} \int_0^{\pi}\frac{-4 \alpha G_F^2(m_{\ell}-2E_{\ell^{\prime}})^5}{ (4\pi)^5 m_{\ell}^2 E_{\ell^{\prime}}} \left(m_{\ell} +2 E_{\ell^{\prime}}\right)\cos{\theta_{\ell'}}d\theta_{\ell^{\prime}} d\phi=0.
  \end{split}
 \end{equation}
 Doing this, it is straightforward that the difference between Dirac and Majorana cases vanishes once the angular integration is made. We emphasize that this difference will vanish in any other selected frame, doing the angular integration properly, as derived in eq.(\ref{eqn:angulargood}).

Specifically, for subsequent discussions, if we work in the system where the neutrinos define the $x$-axis ($\theta_{\nu}=\pi/2$), as done in the KMS method, in such a way that $\cos{\theta}=\cos{\phi}\sin{\theta_{\ell'}}$, the difference between Dirac and Majorana cases after the angular integration is:
 \begin{equation}
   \begin{split}
    &\int_0^{2\pi} \int_0^{\pi}\left( d\Gamma_{\nu\nu}^D|_{b2b}-d\Gamma_{\nu\nu}^M|_{b2b}\right)d\theta_{\ell^{\prime}} d\phi\\
    &=\int_0^{2\pi} \int_0^{\pi}\frac{4 \alpha G_F^2(m_{\ell}-2E_{\ell^{\prime}})^5}{ (4\pi)^5 m_{\ell}^2 E_{\ell^{\prime}}} \left(m_{\ell} +2 E_{\ell^{\prime}}\right)\cos{\phi}\sin{\theta_{\ell'}}d\theta_{\ell^{\prime}} d\phi=0.
  \end{split}
 \end{equation}

Again, the difference between Dirac and Majorana cases vanishes, consistently with the last computation, since the physics must not depend on the selected reference frame we are working on. Here we anticipate that the main discrepancy with the KMS method is that their analysis sets $\phi=0$, where in this example it is clear that setting $\phi=0$ would lead to different results depending on the selected system, which makes no physical sense. This reason, along with several other arguments, will lead us to conclude that $\phi$ is not fixed by any kinematic condition and must be integrated over its entire range, as we will discuss in detail later.
 
\begin{figure}[tbp]
\begin{subfigure}[]{0.5\textwidth}
   \includegraphics[width=1\linewidth]{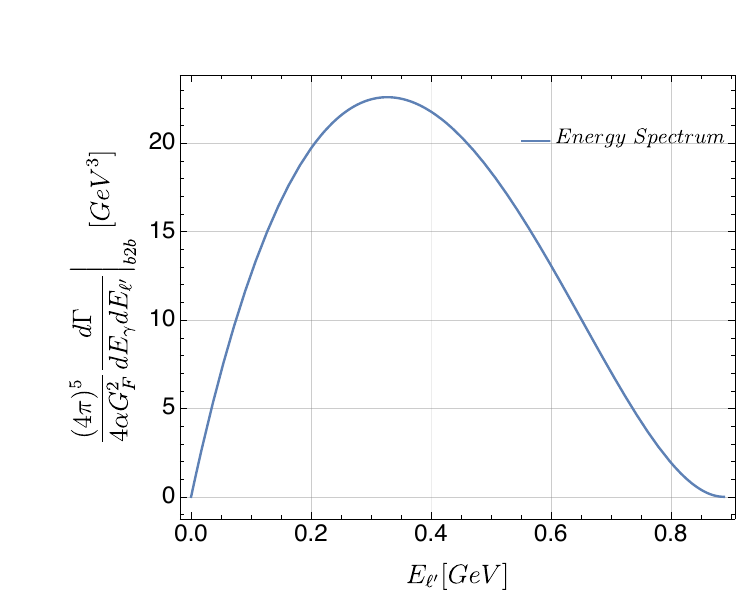}
   \caption{\scriptsize{Charged lepton-photon energy spectrum.}}
   \label{fig: lepgam}
\end{subfigure}
\begin{subfigure}[]{0.5\textwidth}
   \includegraphics[width=1\linewidth]{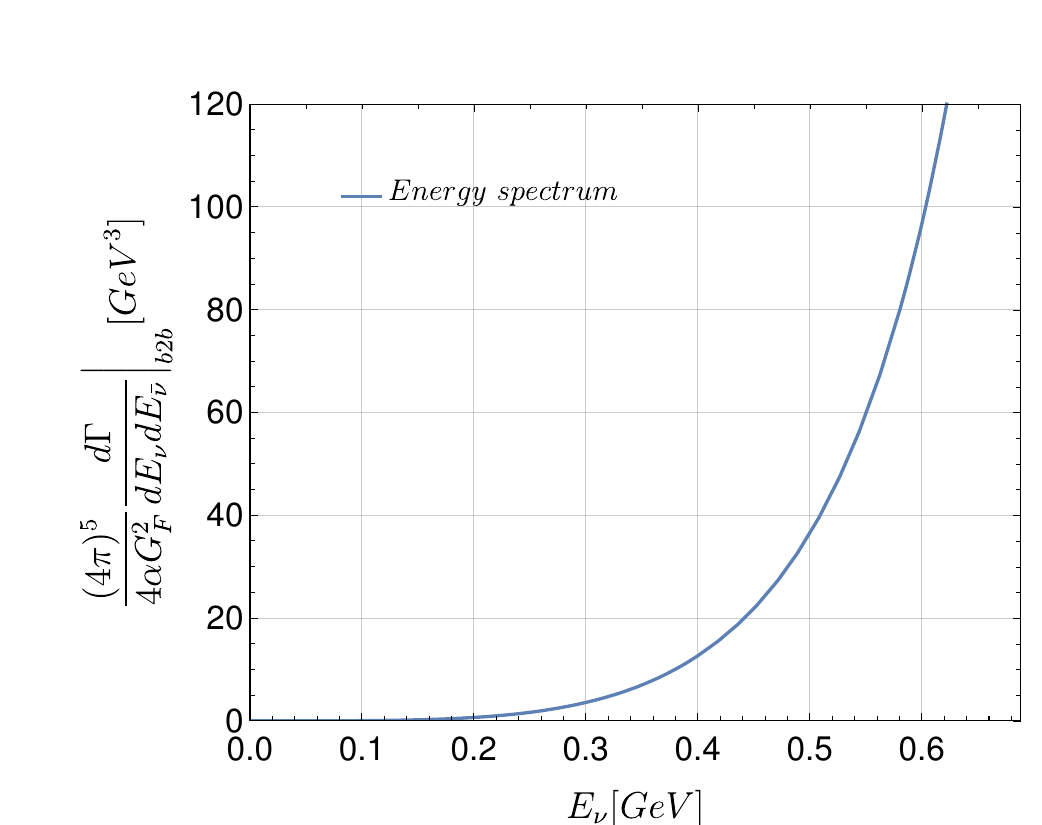}
   \caption{\scriptsize{Neutrinos energy spectrum.}}
   \label{fig:nuantinu} 
\end{subfigure}
 \caption{Energy spectra for charged lepton-photon and neutrinos in $\tau\to\ell'\nu\bar{\nu}\gamma$ decays, restricted to the $b2b$ case. They are identical for Dirac and Majorana cases.}
 \label{fig:energyplot}
 \end{figure}

Actually, we can do the same computation for the $\ell'-\gamma$ decay rate (since, even if in the $b2b$ configuration we can relate $E_{\ell'}$ and $E_{\nu}$, the neutrinos and electron-photon distributions are not the same in this kinematic scenario, as shown in eqs.(\ref{neutrinospectrum}) and (\ref{ellgamaspectrum})) and plot the specific energy distribution for the $\ell'-\gamma$ and $\nu-\Bar{\nu}$ pairs, integrating over the inaccessible angle $\theta$; this is shown in figure \ref{fig:energyplot} (just for a $\tau$ decay, for simplicity).
 As we pointed out before, both distributions are different and the Dirac and Majorana cases are, unfortunately, indistinguishable in each case.
 
 This last point is specially important, since the explanation of the impossibility to distinguish the specific neutrino nature in each distribution is different. For the $\ell'-\gamma$ spectrum, it is a direct application of the DMCT, since we already integrated over all the neutrino variables and we are neglecting neutrino mass contributions. For the $\nu-\Bar{\nu}$ spectrum, one might think that there should be a difference between Dirac and Majorana distributions, since we are keeping information about the neutrinos energies. This would be true if the neutrinos energies were not equal, so the change $E_{\nu}\leftrightarrow E_{\Bar{\nu}}$ could lead to a difference while computing the Majorana neutrino case. Regrettably, in the $b2b$ case, they are exactly the same ($E_{\nu}=E_{\Bar{\nu}}$), so we can not distinguish between Dirac and Majorana nature from this energy variable alone. This statement is precisely the reason why the difference between Dirac and Majorana distributions can not be independent of the angular variable $\theta$.
 

\subsection{Discussions and Consistency Tests}
\label{subsec:3.2}

In this section we elaborate on our results and focus on discussing the main reason for the lack of a difference between Dirac and Majorana nature of neutrinos. This outcome, which may differ from initial expectations based on the motivation provided by the KMS method, deserves careful analysis.
We emphasize that the specific process under study is not the same as the one used in \cite{Kim}, and the Majorana nature affects in a distinct manner the two of them. Nevertheless this will only influence the dynamics and not the kinematics. So, from now on, we will be focusing on the kinematic analysis, which must be the same considering all our previous assumptions.


Particularly, working in the same system as done in the KMS method,  we traced the main difference of this feature with the KMS approach and found that it primarily arises from the $\phi$ variable treatment. Therefore we want to delve further into this topic and specify the reasons why this variable should not be fixed in the $b2b$ configuration. Additionally, we will outline the results that would be obtained if we do so in our specific process.

It is suggested in the KMS method that $\phi=0$ is a condition fixed by the $b2b$ kinematics. As we mentioned, the $b2b$ scenario is obtained by applying the condition $\Vec{p}_2=-\Vec{p}_{3}$, or equivalently $\Vec{p}_{4}=-\Vec{p}_{5}$. These restrictions affect three of the five phase-space kinematic independent variables as follows: $E_{\nu}=E_{\bar{\nu}}$ and $\Theta_{\nu\bar{\nu}}=\pi$. Therefore, the remaining two angular variables must run over all their possible configurations, meaning that we must not set $\phi$ to any specific value.
In other words, the condition $\Vec{p}_2=-\Vec{p}_{3}$ can be achieved for any value of $\phi$ and not just for $\phi=0$. Nevertheless, beyond these qualitative arguments, we did the quantitative derivation of this assertion in Appendix \ref{app:b2b}, where it is shown explicitly that $\phi$ is not fixed at all by the $b2b$ restriction. 

Finally, the last argument given in the KMS approach is that in the $b2b$ configuration the $\nu\bar{\nu}$ and $\ell'\gamma$ systems define a plane (since they are two independent vectors) and thus $\phi=0$. We also clarify this statement in Appendix \ref{app:b2b}, where we fully agree with the $\nu\bar{\nu}$ and $\ell'\gamma$ systems defining a plane, but show that such plane is independent of the $\phi$ value, being $\phi=0$ just an allowed specific configuration. This can be seen directly from the plane equation and applied to any selected system. Then $\phi$ remains an independent variable.

Also, as discussed in previous sections, the physics must not depend on the selected reference frame. We already showed that, at least, in two different systems ($\theta_\nu=0$ and $\theta_\nu=\pi/2$) the difference between Dirac and Majorana distributions vanishes while doing the angular integration properly (not fixing $\phi=0$), which extends to any other. We shall discuss next what happens when assuming the condition $\phi=0$ and will obtain that this difference will be non-vanishing in the system defined by $\theta_\nu=\pi/2$, while it will remain zero in the system where $\theta_\nu=0$, which is clearly a physical contradiction and again another argument against fixing $\phi$ in the $b2b$ kinematic configuration.

Now, for completeness, we will study the possible modifications to our main results in the case that we set $\phi=0$. In the system defined by $\theta_\nu=0$, it is straightforward that the difference in any expression will be just an overall $2\pi$ factor after the angular integration, due to the azimuthal symmetry. Also, for the Dirac-Majorana difference, the result will vanish again due to the $\theta_{\ell'}$ integration.\\
For the following discussion, it is important to compare what happens now in the KMS system ($\theta_\nu=\pi/2$). Replacing the $\phi=0$ condition into eq.(\ref{eq:CommentBelow}), where now $\cos{\theta}=\cos{\phi}\sin{\theta_{\ell'}}=\sin{\theta_{\ell'}}$, the difference between Dirac and Majorana cases is precisely:
 \begin{equation}
 \label{eqn:difdm2}
     d\Gamma_{\nu\nu}^D|_{b2b}-d\Gamma_{\nu\nu}^M|_{b2b}=\frac{4 \alpha G_F^2(m_{\ell}-2E_{\ell^{\prime}})^5}{ (4\pi)^5 m_{\ell}^2 E_{\ell^{\prime}}} \left(m_{\ell} +2 E_{\ell^{\prime}}\right)\sin{\theta_{\ell'}},
 \end{equation} 
 and will be non-zero after the angular integration, as we shall see next, which is a physical contradiction as we just emphasized.
 
\begin{figure}[tbp]
\centering
\begin{subfigure}[]{0.496\textwidth}
   \includegraphics[width=1\linewidth]{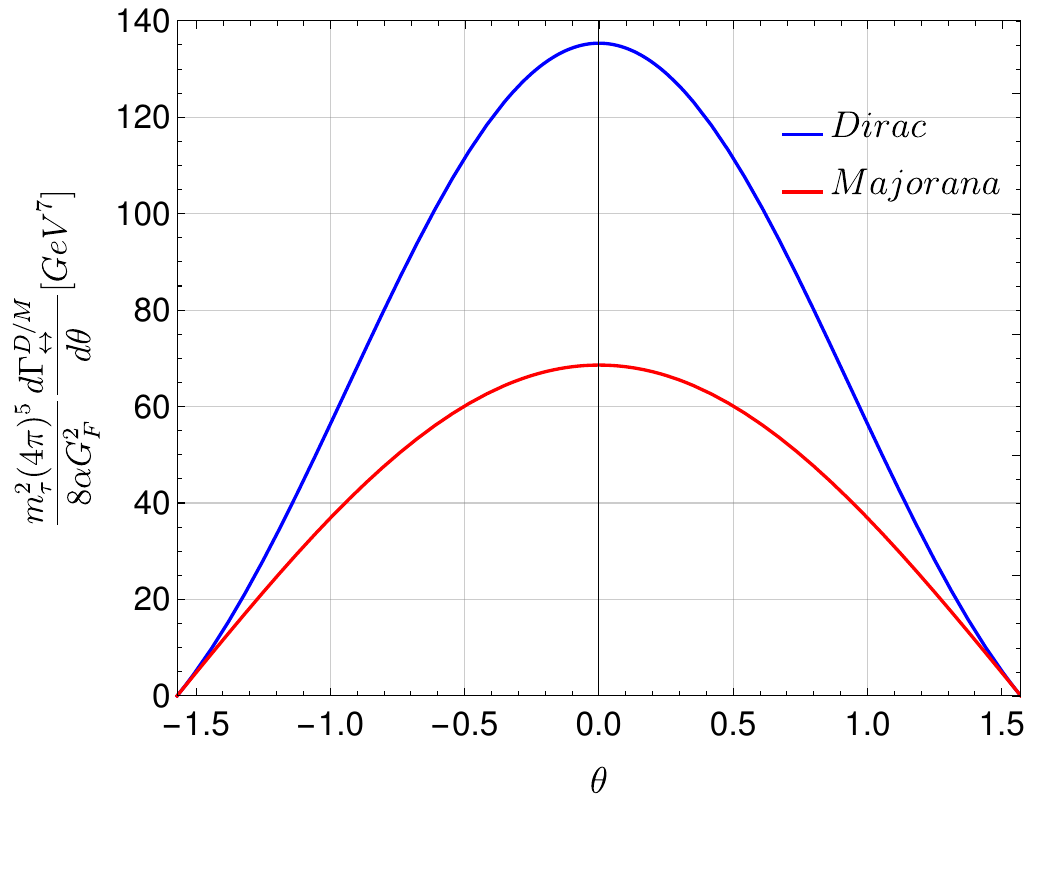}
   \caption{\scriptsize{Comparison of angular distributions between Dirac and Majorana cases in the $b2b$ scenario ($\ell=\tau$).}}
   \label{fig: angular1}
\end{subfigure}
\begin{subfigure}[]{0.496\textwidth}
   \includegraphics[width=1\linewidth]{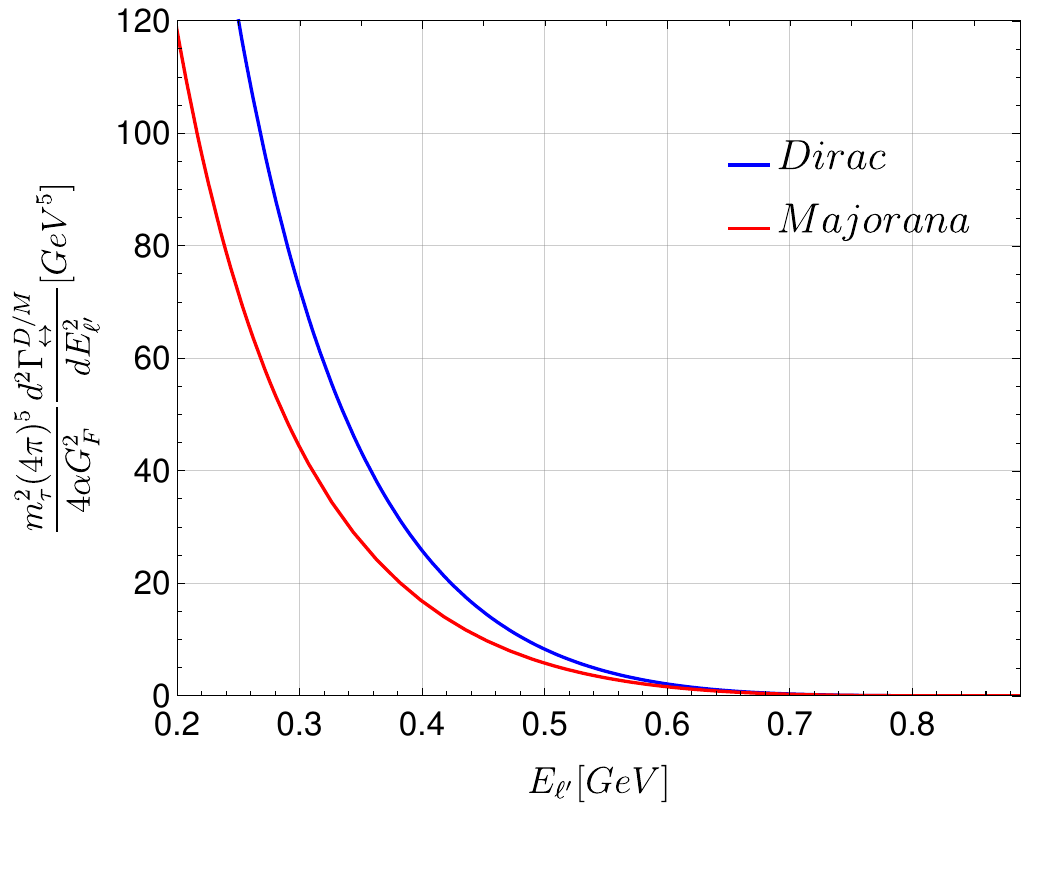}
   \caption{\scriptsize{Comparison of energy distributions between Dirac and Majorana cases in the $b2b$ scenario ($\ell=\tau$).}}
   \label{fig:energy1} 
\end{subfigure}
\begin{subfigure}[]{0.496\textwidth}
   \includegraphics[width=1\linewidth]{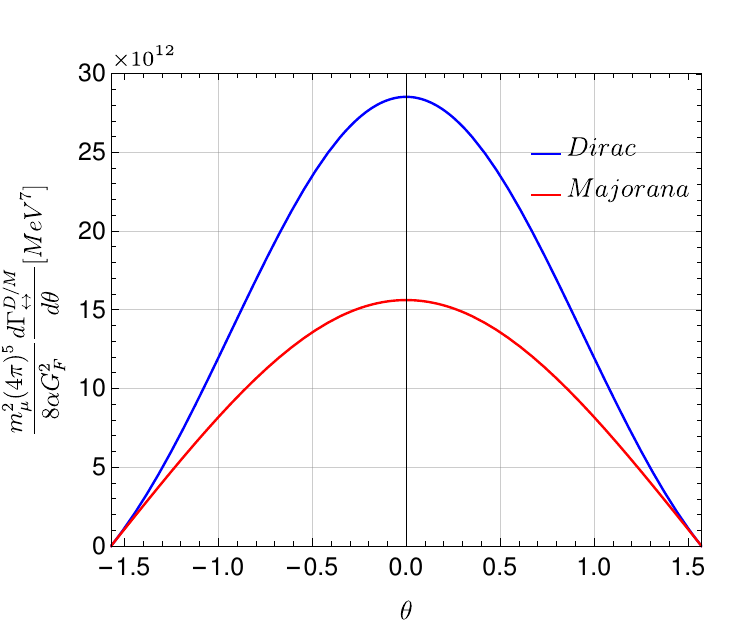}
   \caption{\scriptsize{Comparison of angular distributions between Dirac and Majorana cases in the $b2b$ scenario ($\ell=\mu$).}}
   \label{fig: angular2}
\end{subfigure}
\begin{subfigure}[]{0.496\textwidth}
   \includegraphics[width=1\linewidth]{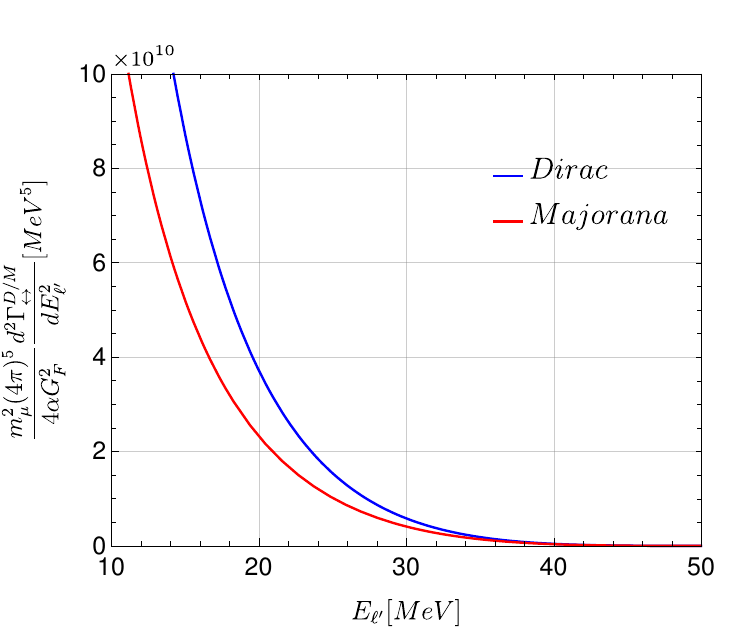}
   \caption{\scriptsize{Comparison of energy distributions between Dirac and Majorana cases in the $b2b$ scenario ($\ell=\mu$).}}
   \label{fig:energy2} 
\end{subfigure}
\caption{\label{fig: energyspec1} Comparison of Dirac and Majorana distributions.}
\label{fig:real}
\end{figure}

 For a complete discussion of this case ($\phi$=0), we can now integrate over the energy range and get the angular distribution shown in figure \ref{fig: angular1} (\ref{fig: angular2}) for the tau (muon) decay, where the difference between Dirac and Majorana cases is obvious. Again, for the experimental observable we must integrate over the inaccessible angle $\theta$ ($\cos{\theta}=\sin{\theta_{\ell'}}$), to get the final neutrinos energy distributions for Dirac and Majorana cases:
  \begin{equation}
     \int d\Gamma_{\nu\nu}^D|_{b2b}\,  d\cos{\theta_{\ell'}}=\frac{4 \alpha G_F^2}{(4\pi)^5 m_{\ell}^2}\frac{(m_{\ell}-2E_{\ell^{\prime}})^4}{ E_{\ell^{\prime}}}\left[\left(\frac{20}{3}-2\pi \right)E_{\ell^{\prime}}^2+\frac{1}{2}(\pi+4)m_{\ell}^2 \right],
     \label{DiracDW}
 \end{equation}
   \begin{equation}
     \int d\Gamma_{\nu\nu}^M|_{b2b}\,  d\cos{\theta_{\ell'}}=\frac{4 \alpha G_F^2}{(4\pi)^5 m_{\ell}^2}\frac{(m_{\ell}-2E_{\ell^{\prime}})^4}{ E_{\ell^{\prime}}}\left[\frac{20}{3}E_{\ell^{\prime}}^2+2m_{\ell}^2 \right].\label{MajDW}
 \end{equation}
 
 This final energy distribution, obtained after integrating over $\cos{\theta_{\ell'}}$, is shown in figure \ref{fig:energy1} (\ref{fig:energy2}) for the tau (muon) decay, considering the Dirac and Majorana cases, with a clear difference between both of them, given explicitly by the following expression:
 \begin{equation}
  \int (d\Gamma_{\nu\nu}^D|_{b2b}-d\Gamma_{\nu\nu}^M|_{b2b})\,  d\cos{\theta_{\ell'}}=\frac{2 \alpha G_F^2}{(4\pi)^5 m_{\ell}^2}\frac{(m_{\ell}-2E_{\ell^{\prime}})^5}{E_{\ell^{\prime}}}\pi\left[m_{\ell}+2 E_{\ell^{\prime}}\right].
 \end{equation}
 Therefore, by setting $\phi=0$, it would be possible in principle to distinguish between Dirac and Majorana nature of neutrinos measuring the final energy distribution of the $b2b$ configuration. Then, this could be a really promising result, motivating its search in current and future experiments, being an attractive alternative process to the one studied in \cite{Kim}, due to its larger branching ratio (BR)\footnote{Unfortunately, if we consider the full angular dependence (without setting $\phi=0$) as we just discussed, the difference between Dirac and Majorana cases vanishes once the angular integration is made, consistently with the result in the last section.}.\\
Finally, focusing on this last statement, we would also like to clarify the estimation process of the BR as done in the KMS method, which could lead to confusion for the $b2b$ case. This is done in detail in Appendix \ref{app:br}, where we find a BR for the $b2b$ case of the following order:
\begin{equation}
\begin{split}
&\mathcal{BR}(\ell=\tau)_{b2b}=\left(\Gamma\right)_{b2b}/\Gamma_{\tau}\approx 1.68\times 10^{-10},\\
&\mathcal{BR}(\ell=\mu)_{b2b}=\left(\Gamma\right)_{b2b}/\Gamma_{\mu}\approx 1.34\times 10^{-10},
\end{split}
\end{equation}
being too suppressed, as expected for such a specific kinematic configuration. \\

In conclusion, once the appropriate treatment of the phase-space in the $b2b$ configuration is clarified, our approach remains consistent with all the tests carried out and allows a clearer interpretation of the results, leading to the fact that there is no difference produced by the Dirac or Majorana nature of neutrinos in $\ell\to\ell'\nu\bar{\nu}\gamma$, independently of the neutrino mass, once the inaccessible neutrino variables are integrated out. Recalling our comments above eq.~(\ref{eq:CommentAbove}) and below eq.~(\ref{eq:CommentBelow}), any difference between these two cases will be heavily suppressed by squared sterile neutrino masses and mixings, likely preventing their soon observation. We remark that we were neglecting possible neutrino non-standard interactions. It remains to be studied how much these can enhance the sensitivity to the neutrino nature~\cite{Kim:2022xjg}, through $\mathcal{BR}(\ell=\mu,\,\tau)_{b2b}$, while still being consistent with all other constraints.

\section{Summary and Conclusions}
\label{sec:4}
In this work we have studied the radiative leptonic lepton-decay $\ell\to \ell'\overline\nu \nu \gamma$, in the neutrinos mass basis. We have developed our own approach, inspired in the method put forward in \cite{Kim}, extending its application to final-state neutrinos of different flavour, in order to distinguish between Dirac and Majorana neutrinos; we have computed its matrix element for the $b2b$ configuration in the decaying lepton rest frame for both cases.

Presumably, this $b2b$ kinematic scenario avoids the constraint imposed by the ‘practical Dirac-Majorana confusion theorem’ (DMCT)\cite{Kayser} leading to striking differences between the Dirac and Majorana cases, that are not proportional to tiny neutrino masses. Instead, we found that there is no difference between Dirac and Majorana distribution in $\ell\to \ell'\overline\nu \nu \gamma$ once the inaccessible neutrino angle is integrated out.

We discussed in detail the angular treatment, with quantitative and qualitative arguments favoring our conclusions. This turns out to be crucial, since its inaccurate interpretation could lead to very attractive results, even observable in current and near future experiments. However, after careful study, this leads to no difference between Dirac and Majorana energy distributions, for the process under consideration, remaining consistent with all the tests done.

Finally, we wish to emphasize that the idea proposed by \cite{Kim} is very appealing in order to avoid the DMCT. This fact highlights the necessity to study other types of decays and specific kinematic scenarios within this approach, where angular or energy dependencies could lead to a non-zero difference between the Dirac and Majorana distributions, hopefully observable in current or forthcoming experiments. 

\section*{Acknowledgements}
The authors are indebted to Gabriel López Castro for many enlightening discussions during the completion of this work. J.~M.~M\'arquez and D.~Portillo acknowledge useful correspondence with C. S. Kim and Dibyakrupa Sahoo. J. M. M. and D. P. S. are thankful to Conacyt funding their Ph. D. We also acknowledge the support from Ciencia de Frontera Conacyt (México) project No. 428218. P. R.  was partly funded
by ‘Cátedras Marcos Moshinsky’ (Fundación Marcos
Moshinsky) and ‘Paradigmas y Controversias de la Ciencia 2022’ (project number 319395, Conacyt) during this research.

\appendix

\section{Phase space definition}
\label{app:phasespace}

First of all, in order to avoid confusion, we define our differential rate notation for the process $\ell(p_1)\to\bar{\nu}(p_2)\nu(p_3)\ell^{\prime}(p_4)\gamma(p_5)$. Upon the definition of the following five independent variables of the system, in the frame typically used in a four-body decay process (as can be seen in Fig.[4] of ref.\cite{Kim}), we have:
\begin{itemize}
    \item $s_{\nu\bar{\nu}}\equiv (p_2+p_3)^2$ and $s_{\ell^{\prime}\gamma}\equiv (p_4+p_5)^2$, the invariant masses of the $\nu-\bar{\nu}$ and $\ell^{\prime}-\gamma$ systems, respectively.
    \item $\theta_{\nu}$ ($\theta_{\ell^{\prime}}$), the polar angle between the three-momentum of $\bar{\nu}$ ($\ell^{\prime}$), in the center-of-momentum frame of the $\bar{\nu}-\nu$ ($\ell^{\prime}-\gamma$) pair, and the flight direction of the $\nu-\bar{\nu}$ ($\ell^{\prime}-\gamma$) system in the rest frame of $\ell$.
    \item $\phi$, the azimuthal angle described between the two planes defined by the $\nu-\bar{\nu}$ and $\ell^{\prime}-\gamma$ systems, in the rest frame of $\ell$.
\end{itemize}
We can write the differential decay width as 
\begin{equation}
    \frac{d\Gamma}{d s_{\nu\bar{\nu}} d s_{\ell^{\prime}\gamma} d \cos{\theta_{\nu} }d\cos{\theta_{\ell^{\prime}}}d \phi}= \frac{X \beta_{\nu}\beta_{\ell^{\prime}}}{(4\pi)^6 m_{\ell} \sqrt{s_{\nu\bar{\nu}}s_{\ell^{\prime}\gamma}}}|\overline{\mathcal{M}}|^2 ,
    \label{d5gamma}
\end{equation}
with
\begin{equation}
    X\equiv\frac{\sqrt{\lambda(m_{\ell}^2,s_{\nu\bar{\nu}},s_{\ell^{\prime}\gamma})}}{2m_{\ell}},\quad \beta_{\nu}\equiv\sqrt{\frac{\lambda(s_{\nu\bar{\nu}},m_\nu^2,m_{\bar{\nu}}^2)}{4s_{\nu\bar{\nu}}}},\quad \beta_{\ell^{\prime}}\equiv\sqrt{\frac{\lambda(s_{\ell^{\prime}\gamma},m_{\ell^{\prime}}^2,0)}{4s_{\ell^{\prime}\gamma}}} ,\label{betadef}
\end{equation}
where $X$ is the magnitude of three-momentum of $\nu-\bar{\nu}$ or $\ell^{\prime}-\gamma$ system in the rest frame of $\ell$, while $\beta_{\nu}$ ($\beta_{\ell^{\prime}}$) refers to the magnitude of three-momentum of the $\nu$  ($\ell^{\prime}$) in the center-of-momentum frame of the $\nu\bar{\nu}$ ($\ell^{\prime}\gamma$) pair. 

It is convenient to rewrite the differential width in terms of $E_{\nu}$ and $E_{\bar{\nu}}$ in order to obtain the energy spectrum of the neutrinos. Thus, we can do the variables change:
 \begin{equation}
    ds_{\nu\bar{\nu}}ds_{\ell^{\prime}\gamma}d\cos{ \theta_\nu}=-\frac{4m_{\ell}\sqrt{s_{\nu\bar{\nu}}}}{X\beta_{\nu}}\sqrt{(E_{\nu}^2-m_{\nu}^2)(E_{\bar{\nu}}^2-m_{\bar{\nu}}^2)}dE_{\nu} dE_{\bar{\nu}} d \cos\Theta_{\nu\bar{\nu}},
\end{equation}
where $\Theta_{\nu\bar{\nu}}$ is the angle between the three-momenta of both neutrinos, in the rest frame of $\ell$. Also, we can obtain the energy spectrum for the final charged lepton and the photon with the following change of variables:
\begin{equation}
        ds_{\nu\bar{\nu}}ds_{\ell^{\prime}\gamma}d \cos{\theta_{\ell'}}=-\frac{4m_{\ell}\sqrt{s_{\ell^{\prime}\gamma}}}{X\beta_{\ell^{\prime}}}E_{\gamma}\sqrt{(E_{\ell^{\prime}}^2-m_{\ell^{\prime}}^2)}dE_{\ell^{\prime}} dE_{\gamma} d \cos\Theta_{\ell^{\prime}\gamma},
\end{equation}
where $\Theta_{\ell^{\prime}\gamma}$ is the angle between $\ell^{\prime}$ and $\gamma$, in the rest frame of $\ell$.

Finally, neglecting all the final-state masses, as a good approximation, we get for the neutrinos differential decay distribution:
\begin{equation}
    \frac{d\Gamma^{D,M}}{dE_{\nu}dE_{\Bar{\nu}}d\cos\Theta_{\nu\bar{\nu}}d\cos\theta_{\ell^{\prime}}d\phi}= \frac{2}{m_{\ell}(4\pi)^6}\frac{E_{\nu}E_{\Bar{\nu}}E_{\ell'}}{E_{\gamma}}\frac{1}{\epsilon}\sum_{j,k}|\overline{\mathcal{M}^{D,M}}|^2,
\end{equation}
where $\epsilon=1 (2)$ for Dirac (Majorana) neutrinos. Here $E_{\ell'}$ and $E_{\gamma}$ must be written in terms of $E_{\Bar{\nu}}$ and $E_{\nu}$, according to the energy-momentum conservation law. Note also that we are taking into account all the possible neutrino mass final states and the sum extends over all energetically allowed neutrino pairs.
 The $1/2$ factor that appears in the Majorana case has two different origins. For the $k=j$ case, it is due to the presence of indistinguishable fermions in the final state; while for $k\neq j$, it arises because of double counting, since the sum $\sum\limits_{j,k}$ is not restricted to  $j\leq k$.

Meanwhile, for the differential decay distribution involving the charged lepton and photon energies we obtain:
\begin{equation}
    \frac{d\Gamma^{D,M}}{dE_{\ell'}dE_{\gamma}d\cos{\Theta_{\ell^{\prime}\gamma}}d\cos{\theta_{\nu}}d\phi}= \frac{2}{m_{\ell}(4\pi)^6}\frac{E_{\nu}E_{\ell'}E_{\gamma}}{E_{\Bar{\nu}}}\frac{1}{\epsilon}\sum_{j,k}|\overline{\mathcal{M}^{D,M}}|^2,
    \label{diffwidth}
\end{equation}
where $E_{\Bar{\nu}}$ and $E_{\nu}$ must be written in terms of $E_{\ell'}$ and $E_{\gamma}$ according to the energy-momentum conservation law. These, in principle, are two different spectra and will be so in any specific kinematic configuration.

\section{Back-to-back configuration}
\label{app:b2b}
As written above, it is convenient to describe our phase space variables in the rest frame of the decaying particle to avoid any confusion. First, we are going to denote $p_i'$ as the momentum of the $i-$particle in the corresponding center of mass frame for the relevant particle pair ($\nu-\bar{\nu}$ or $\ell'-\gamma$) and $p_i$ the corresponding momentum in the rest frame of the decaying particle. Now, following Fig.[4] of ref.\cite{Kim}, both momenta, $p_i'$ and $p_i$, are related by a Lorentz boost in the $\hat{z}$ direction. 
We define our boost in the $\hat{z}$ direction for the 4-momentum $p'_{\nu}$ and $p'_{\bar{\nu}}$ as follows:
\begin{equation}
    \Lambda^{\mu}{}_{\nu}=\left(\begin{tabular}{cccc}
       $\sqrt{1+\frac{X^2}{s_{\nu\bar{\nu}}}}$  & 0 & 0 & $\frac{X}{\sqrt{s_{\nu\bar{\nu}}}}$  \\
        0 & 1 & 0 & 0\\
        0 & 0& 1 & 0 \\
        $\frac{X}{\sqrt{s_{\nu\bar{\nu}}}}$ & 0 & 0 & $\sqrt{1+\frac{X^2}{s_{\nu\bar{\nu}}}}$
    \end{tabular}\right).\label{boost1}
\end{equation}
For $p'_{4}$ and $p'_5$ we use the Lorentz transformation:
\begin{equation}
    \Lambda'^{\mu}{}_{\nu}=\left(\begin{tabular}{cccc}
       $\sqrt{1+\frac{X^2}{s_{\ell'\gamma}}}$  & 0 & 0 & $-\frac{X}{\sqrt{s_{\ell'\gamma}}}$  \\
        0 & 1 & 0 & 0\\
        0 & 0& 1 & 0 \\
        $-\frac{X}{\sqrt{s_{\ell'\gamma}}}$ & 0 & 0 & $\sqrt{1+\frac{X^2}{s_{\ell'\gamma}}}$
    \end{tabular}\right),\label{boost2}
\end{equation}
where we employ the definition of $X$, $s_{\nu\bar{\nu}}$ and $s_{\ell'\gamma}$ from the previous appendix. In general, we can write the corresponding 4-momentum in the rest frame of $\ell$ as follows:

\begin{equation}
\begin{split}
    p_2&=\left(\begin{tabular}{cccc}
       $\sqrt{1+\frac{X^2}{s_{\nu\bar{\nu}}}}$  & 0 & 0 & $\frac{X}{\sqrt{s_{\nu\bar{\nu}}}}$  \\
        0 & 1 & 0 & 0\\
        0 & 0& 1 & 0 \\
        $\frac{X}{\sqrt{s_{\nu\bar{\nu}}}}$ & 0 & 0 & $\sqrt{1+\frac{X^2}{s_{\nu\bar{\nu}}}}$
    \end{tabular}\right) \left(\begin{tabular}{c}
         $\frac{\sqrt{s_{\nu\bar{\nu}}}}{2}$ \\
         $\beta_{\nu}\sin\theta_{\nu}\cos\phi$\\
         $\beta_{\nu}\sin\theta_{\nu}\sin\phi$\\
         $\beta_{\nu}\cos\theta_{\nu}$\\
    \end{tabular}\right)\\
    &=\left(\begin{tabular}{c}
         $\frac{\sqrt{s_{\nu\bar{\nu}+X^2}}}{2}+\frac{X}{\sqrt{s_{\nu\bar{\nu}}}}\beta_{\nu}\cos{\theta_{\nu}}$\\
         $\beta_{\nu}\sin\theta_{\nu}\cos\phi$\\
         $\beta_{\nu}\sin\theta_{\nu}\sin\phi$\\
         $\frac{X}{2}+\sqrt{1+\frac{X^2}{s_{\nu\bar{\nu}}}}\beta_{\nu}\cos\theta_{\nu}$\\
    \end{tabular}\right),
    \end{split}
\end{equation}

\begin{equation}
\begin{split}
    p_3&=\left(\begin{tabular}{cccc}
       $\sqrt{1+\frac{X^2}{s_{\nu\bar{\nu}}}}$  & 0 & 0 & $\frac{X}{\sqrt{s_{\nu\bar{\nu}}}}$  \\
        0 & 1 & 0 & 0\\
        0 & 0& 1 & 0 \\
        $\frac{X}{\sqrt{s_{\nu\bar{\nu}}}}$ & 0 & 0 & $\sqrt{1+\frac{X^2}{s_{\nu\bar{\nu}}}}$
    \end{tabular}\right) \left(\begin{tabular}{c}
         $\frac{\sqrt{s_{\nu\bar{\nu}}}}{2}$ \\
         $-\beta_{\nu}\sin\theta_{\nu}\cos\phi$\\
         $-\beta_{\nu}\sin\theta_{\nu}\sin\phi$\\
         $-\beta_{\nu}\cos\theta_{\nu}$\\
    \end{tabular}\right)\\
    &=\left(\begin{tabular}{c}
         $\frac{\sqrt{s_{\nu\bar{\nu}+X^2}}}{2}-\frac{X}{\sqrt{s_{\nu\bar{\nu}}}}\beta_{\nu}\cos{\theta_{\nu}}$\\
         $-\beta_{\nu}\sin\theta_{\nu}\cos\phi$\\
         $-\beta_{\nu}\sin\theta_{\nu}\sin\phi$\\
         $\frac{X}{2}-\sqrt{1+\frac{X^2}{s_{\nu\bar{\nu}}}}\beta_{\nu}\cos\theta_{\nu}$\\
    \end{tabular}\right).
    \end{split}
\end{equation}
\begin{equation}
\begin{split}
    p_{4}&=\left(\begin{tabular}{cccc}
       $\sqrt{1+\frac{X^2}{s_{\ell'\gamma}}}$  & 0 & 0 & $-\frac{X}{\sqrt{s_{\ell'\gamma}}}$  \\
        0 & 1 & 0 & 0\\
        0 & 0& 1 & 0 \\
        $-\frac{X}{\sqrt{s_{\ell'\gamma}}}$ & 0 & 0 & $\sqrt{1+\frac{X^2}{s_{\ell'\gamma}}}$
    \end{tabular}\right)\left(\begin{tabular}{c}
          $\frac{\sqrt{s_{\ell'\gamma}}}{2}$\\
          $\beta_{\ell'}\sin\theta_{\ell'}$\\
          0\\
          $-\beta_{\ell'}\cos\theta_{\ell'}$\\
    \end{tabular}\right)\\
    &=\left(\begin{tabular}{c}
         $\frac{\sqrt{s_{\ell'\gamma}+X^2}}{2}-\frac{X}{\sqrt{s_{\ell'\gamma}}}\beta_{\ell'}\cos{\theta_{\ell'}}$\\
         $\beta_{\ell'}\sin\theta_{\ell'}$\\
         0\\
         $-\frac{X}{2}-\sqrt{1+\frac{X^2}{s_{\ell'\gamma}}}\beta_{\ell'}\cos\theta_{\ell'}$\\
    \end{tabular}\right),
    \end{split}
\end{equation}
\begin{equation}
\begin{split}
    p_{5}&=\left(\begin{tabular}{cccc}
       $\sqrt{1+\frac{X^2}{s_{\ell'\gamma}}}$  & 0 & 0 & $-\frac{X}{\sqrt{s_{\ell'\gamma}}}$  \\
        0 & 1 & 0 & 0\\
        0 & 0& 1 & 0 \\
        $-\frac{X}{\sqrt{s_{\ell'\gamma}}}$ & 0 & 0 & $\sqrt{1+\frac{X^2}{s_{\ell'\gamma}}}$
    \end{tabular}\right)\left(\begin{tabular}{c}
          $\frac{\sqrt{s_{\ell'\gamma}}}{2}$\\
          $-\beta_{\ell'}\sin\theta_{\ell'}$\\
          0\\
          $\beta_{\ell'}\cos\theta_{\ell'}$\\
    \end{tabular}\right)\\
    &=\left(\begin{tabular}{c}
         $\frac{\sqrt{s_{\ell'\gamma}+X^2}}{2}+\frac{X}{\sqrt{s_{\ell'\gamma}}}\beta_{\ell'}\cos{\theta_{\ell'}}$\\
         $-\beta_{\ell'}\sin\theta_{\ell'}$\\
         0\\
         $-\frac{X}{2}+\sqrt{1+\frac{X^2}{s_{\ell'\gamma}}}\beta_{\ell'}\cos\theta_{\ell'}$\\
    \end{tabular}\right).
    \end{split}
\end{equation}

Finally, we can apply the $b2b$ constraint $\Vec{p}_2=-\Vec{p}_3$ or equivalently, due to energy momentum conservation,  $\vec{p}_4=-\Vec{p}_5$.
In this kinematic scenario it is easy to show that $X=0$, which is consistent with the fact that, in the $b2b$ configuration, the boosts in eqs.(\ref{boost1},\ref{boost2}) are exactly the identity matrix, i.e., the center of mass frame of the $\nu-\bar{\nu}$ and $\ell'-\gamma$ systems coincides with the decaying lepton rest frame. Then, in the $b2b$ case the three-momentum of the final-state particles can be written as follows

\begin{equation}
\Vec{p}_2=\left(\begin{tabular}{c}
$\beta_{\nu} \sin\theta_{\nu} \cos\phi$ \\
$\beta_{\nu} \sin\theta_{\nu} \sin\phi$ \\
$\beta_{\nu} \cos\theta_{\nu}$\\
\end{tabular}\right), \quad \quad
\Vec{p}_3=\left(\begin{tabular}{c}
$- \beta_{\nu} \sin\theta_{\nu} \cos\phi$ \\
$- \beta_{\nu} \sin\theta_{\nu} \sin\phi$ \\
$- \beta_{\nu}\cos\theta_{\nu}$\\
\end{tabular}\right),
\end{equation}
\begin{equation}
\Vec{p}_4=\left(\begin{tabular}{c}
$\beta_{\ell'} \sin\theta_{\ell'}$ \\
0 \\
$-\beta_{\ell'} \cos\theta_{\ell'}$\\
\end{tabular}\right), \quad \quad
\Vec{p}_5=\left(\begin{tabular}{c}
$-\beta_{\ell'} \sin\theta_{\ell'}$ \\
0 \\
$\beta_{\ell'} \cos\theta_{\ell'}$\\
\end{tabular}\right).
\end{equation}

Here it is essential to emphasize that the above equations fulfill the $b2b$ constraint ($\Vec{p}_2=-\Vec{p}_3$ and $\Vec{p}_4=-\Vec{p}_5$) for all possible values of $(\theta_{\ell'}, \theta_\nu, \phi)$ and not only when $\phi=0$. Thus, we showed cleverly that $\phi=0$ is not a constraint imposed by the $b2b$ kinematics, as suggested in \cite{Kim} (see also \cite{Kim2}), and it needs to be integrated over its full range. 

Another important result comes from the definition of the angle $\theta$ that appears in the amplitude, which is the angle between the neutrino and the charged lepton. From that definition, it is straightforward to compute its explicit form, in terms of the angles $\theta_{\ell'}, \theta_\nu$ and $\phi$, for the $b2b$ configuration. Using the above expressions for the three-momentum in the $b2b$ case we obtained: 
\begin{equation}
    \begin{split}
       \cos{\theta}&\equiv\hat{p}_{2}\cdot\hat{p}_4=\sin{\theta_{\ell^{\prime}}}\sin{\theta_{\nu}}\cos{\phi}-\cos{\theta_{\nu}}\cos{\theta_{\ell^{\prime}}},
    \end{split}
    \label{eqn:angulargood}
\end{equation}
that shows the specific dependence of $\theta$ on $\theta_{\nu}$, $\theta_{\ell'}$ and $\phi$. 
Other relations resulting from $\Vec{p}_2=-\Vec{p}_3$ and $\Vec{p}_4=-\Vec{p}_5$ (neglecting the mass of the final-state particles) are

\begin{equation}
    \begin{split}
    & s_{\nu\nu}=4E_\nu^2,\quad s_{\ell'\gamma}=(m_{\ell}-2E_\nu)^2,\\
       &\beta_{\ell'}=\frac{m_{\ell}}{2}-E_\nu ,\quad\beta_{\nu}=E_\nu,\\
       &\Theta_{\nu\bar{\nu}}=\pi, \quad \Theta_{\ell'\gamma}=\pi,
    \end{split}
\end{equation}
taking $E_1=E_2=E_{\nu}$. 

Finally, since in this b2b configuration the $\Vec{p}_2 (-\Vec{p}_3)$ and  $\Vec{p}_4 (-\Vec{p}_5)$ are two independent vectors, they can always span a plane, i.e. they can always form a basis of a two-dimensional space. This argument is used in the KMS method to claim that $\phi=0$. For completeness we work on this subject below and demonstrate that, even it is certainly true that these two vectors span a plane, this condition does not fix $\phi=0$, as we just showed before.  \\

\textbf{How can we describe that plane?}\\
Since in the $\ell$ rest frame, these vectors start from the origin of the coordinate system $(x_0,y_0,z_0)=(0,0,0)$, then the plane spanned by the vectors $\Vec{p}_2$ and  $\Vec{p}_4$ is given by the well-known equation:
\begin{equation}
    (x, y, z)=\lambda \Vec{p}_2 + \nu \Vec{p}_4,
\end{equation}
where $\lambda$ and $\nu$ are just the parameters of the plane-equation $(-\infty<\lambda, \nu<+\infty)$. 

\begin{figure}[tbp]
    \centering
    \begin{tabular}{cc}
        \includegraphics[scale=.35]{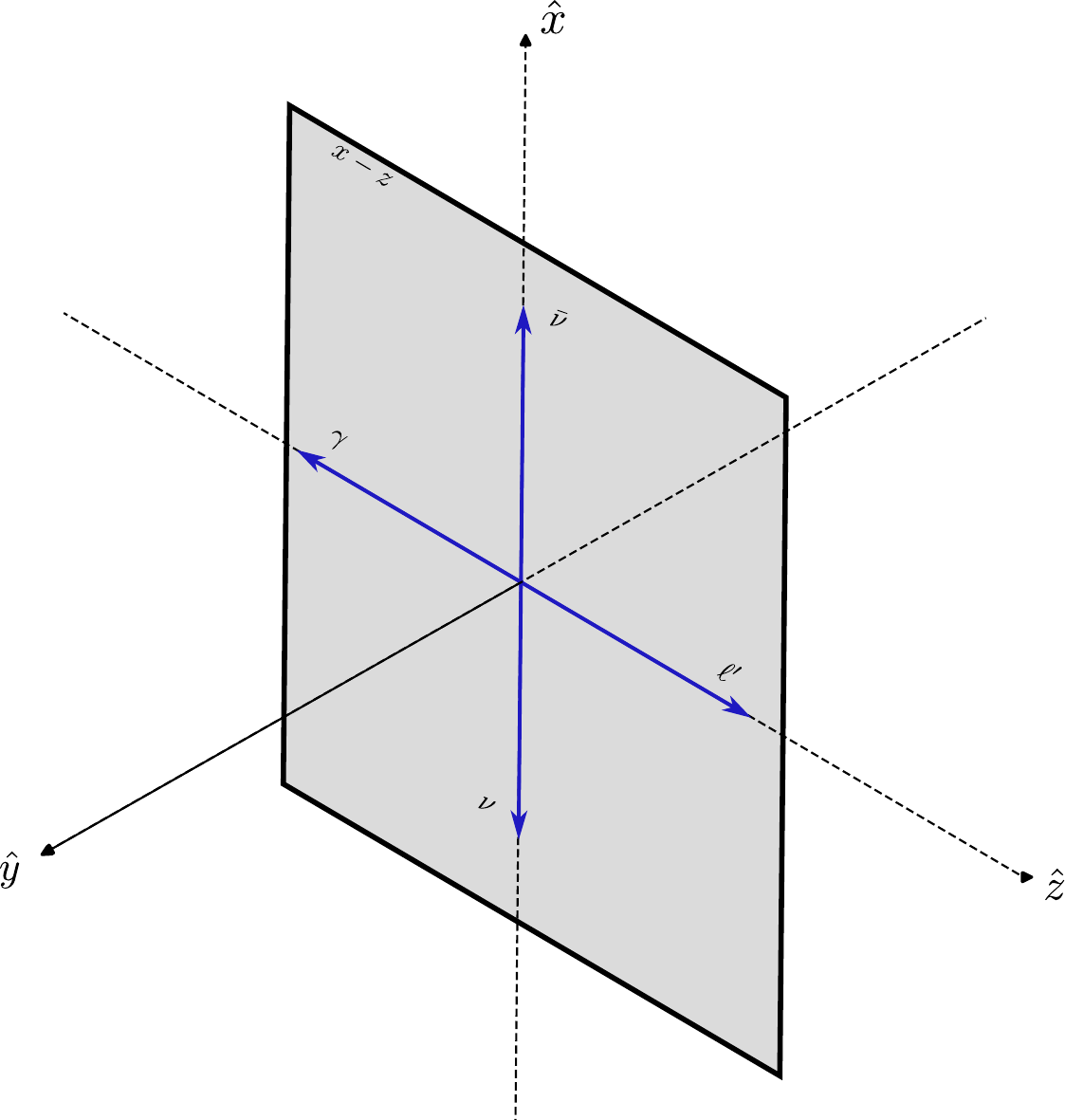} & \includegraphics[scale=.35]{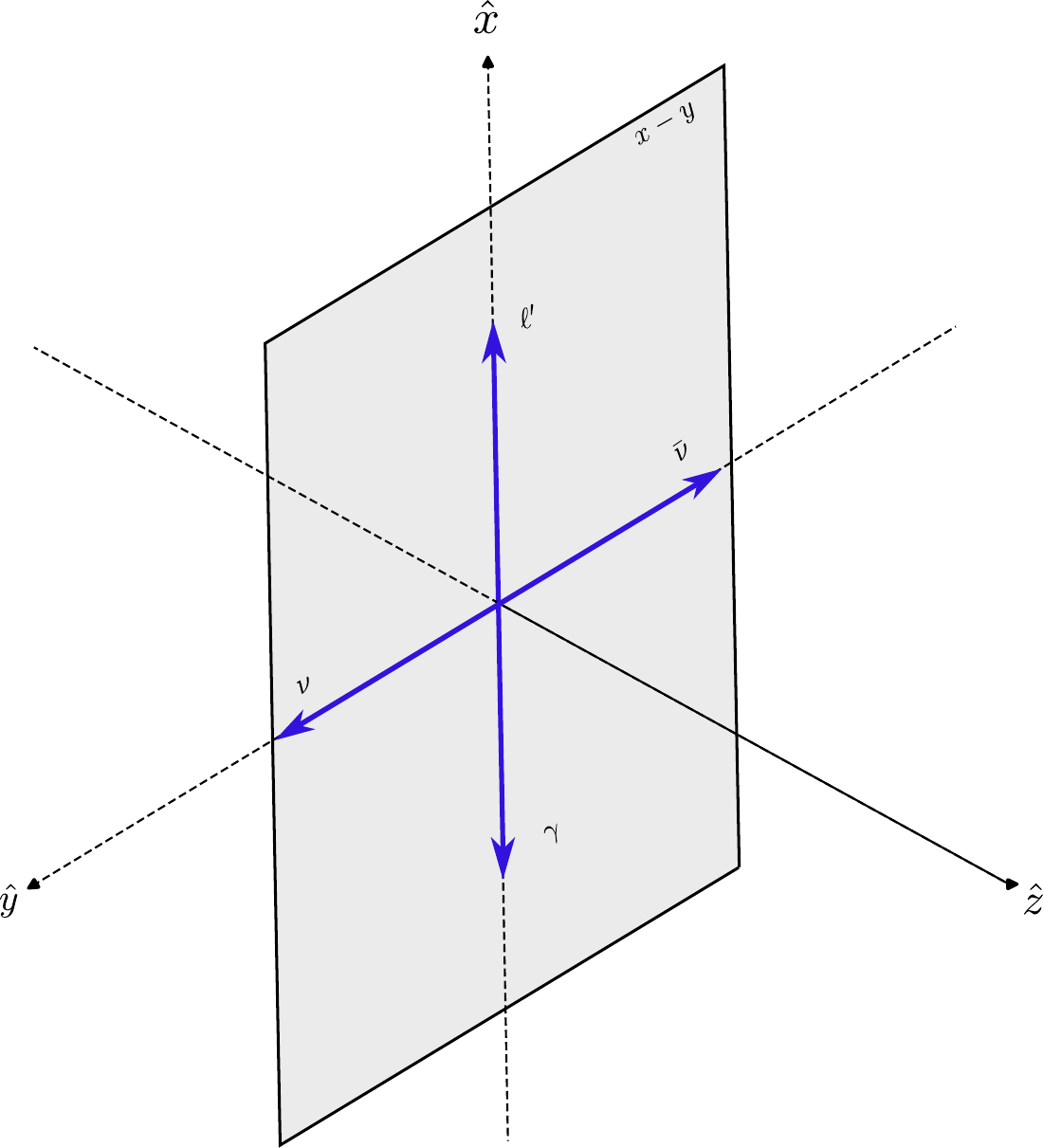}\\ 
        (a) & (b)\\
        $\phi=0$ & $\theta_{\nu}=\pi/2$ \\
    \end{tabular}
    \begin{tabular}{c}
         \includegraphics[scale=.35]{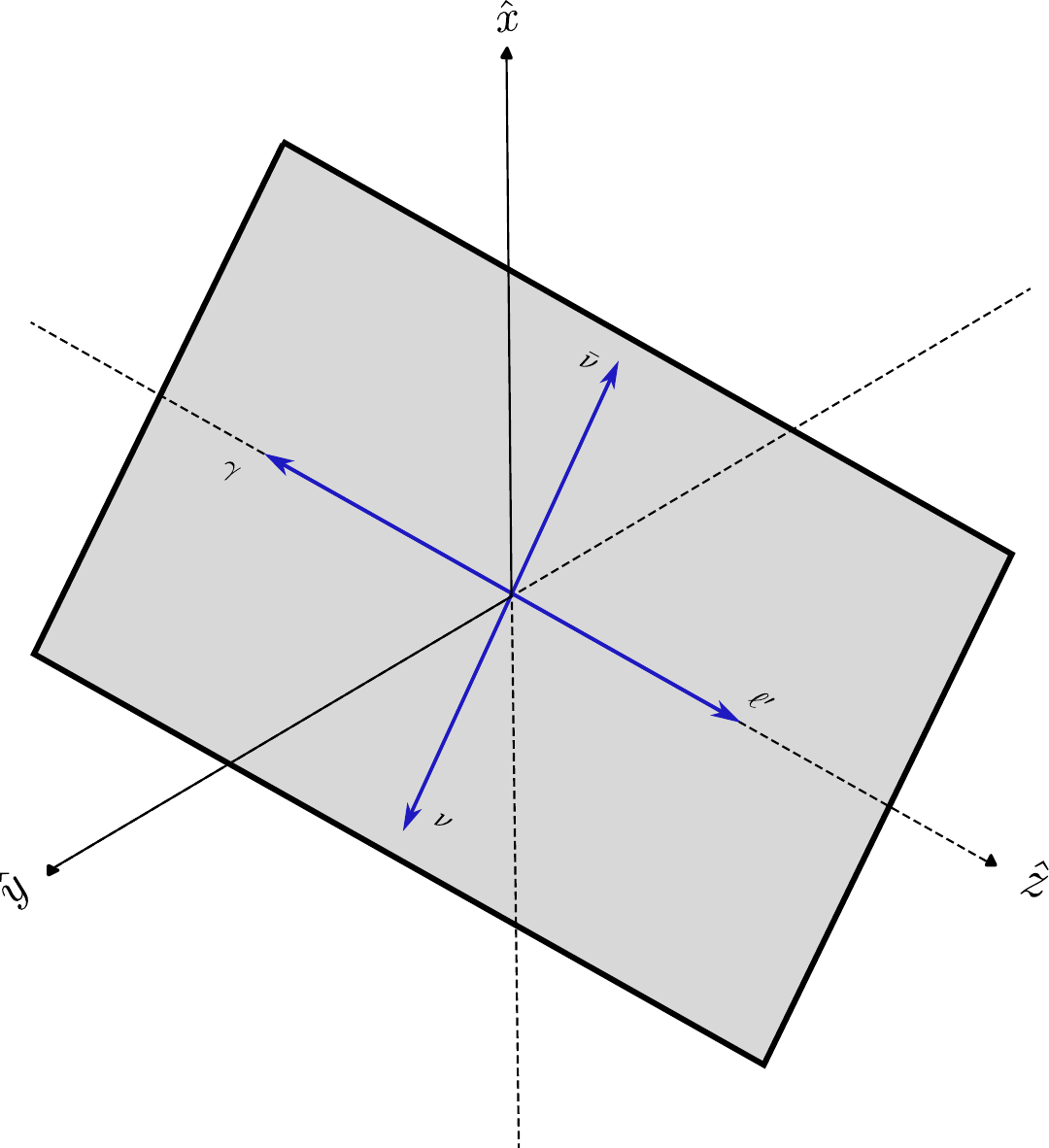}\\
         (c)\\
        $\theta_{\ell'}=0$, $\phi=-\pi/4$\\
    \end{tabular}
    \caption{Planes defined from $\Vec{p}_2$ and $\Vec{p}_4$ in the b2b depending on different values of $\phi$, $\theta_{\ell'}$.}
    \label{fig:planes}
\end{figure}

The following computation can be done in any selected reference frame, but for this specific discussion we keep working in the system where $\theta_{\nu}=\pi/2$ as done in the KMS method, where the plane equation can be put in the general form, giving rise to (after a fast simplification):
\begin{equation}
    (\cos\theta_{\ell'} \sin\phi)x-(\cos\theta_{\ell'} \cos\phi)y+(\sin\theta_{\ell'} \sin\phi)z=0. \label{plane}
\end{equation}
Then it is completely clear that the b2b condition is satisfied for each value of $\phi$ and the corresponding plane spanned by the vectors $\Vec{p}_2$ and  $\Vec{p}_4$ is given by eq.(\ref{plane}). This reaffirms that $\phi=0$ is not a condition fixed by the b2b scenario, and instead $\phi$ remains as an independent variable that must be integrated over its full range.\\
To illustrate this plane condition, we show in Fig. \ref{fig:planes} various examples with different $\phi$ and $\theta_{\nu}$:\\

$\bullet$ Diagram (a): For $\phi=0$ we get $y=0$, which means that the plane is precisely the $x-z$ plane, which is in agreement with the KMS method, but is not the only option allowed by the kinematics.\\

$\bullet$ Diagram (b): For $\theta_{\ell'}=\pi/2$ we get $z=0$, which means that the plane is precisely the $x-y$ one, a configuration completely allowed by the kinematics.\\

And you can keep going with all the possible configurations, as diagram (c), etc. All of them are in agreement with the $b2b$ constraints and with the fact that the two final independent vectors span a plane, checking again that $\phi=0$ is just an allowed configuration but not a condition fixed by the $b2b$ kinematics.

\section{Branching ratio computation}
\label{app:br}
First we will compute the BR of this $b2b$ kinematic scenario in a way that might seem correct at first glance, but that -without the right considerations- can lead to erroneous conclusions. Finally we will discuss this problem in detail and the correct way to estimate this observable.

 For a first consistency test, the total BR($\ell\to\ell'\nu
\bar{\nu}
\gamma$) for the general kinematic configuration was computed, being in perfect agreement with those reported by \cite{Fael:2015gua}, giving us a corroboration that our procedure was correct.
  
 Now, as a first attempt, one might be tempted to estimate the $b2b$ BR by integrating over all the energy and angular configurations of the differential decay rates evaluated in this kinematic case. Essentially, for the case $\phi=0$, integrating over the remaining energy dependence of eqs. (\ref{DiracDW}, \ref{MajDW}):
 \begin{equation}
     \mathcal{B}^{\{D,M\}}_{\leftrightarrow}\equiv\frac{1}{\Gamma_{\ell}}\int \left(d\Gamma_{\nu\nu}^{D,M}|_{b2b}\right)dE_{\nu}dE_{\bar{\nu}}d\cos{\theta_{\ell'}}\,.
 \end{equation}
 
Doing this, and cutting off photons below an energy threshold of $10$ MeV in the decaying-lepton rest frame, given by the experimental resolution at Belle \cite{Belle:2017wxw} (achievable at Belle-II \cite{Belle-II:2018jsg}) \footnote{This threshold was $7$ MeV at the Crystal Box experiment, in the search for $\mu\to e\gamma\gamma$ \cite{Bolton:1988af}.}, we obtain:
 
\begin{equation}
\label{eqn:BRt}
\begin{split}
&\mathcal{B}^D_{\leftrightarrow}(\ell=\tau)=\Gamma^{D}_{\leftrightarrow}/\Gamma_{\tau}\approx 4.3\times 10^{-4},\\
&\mathcal{B}^M_{\leftrightarrow}(\ell=\tau)=\Gamma^{M}_{\leftrightarrow}/\Gamma_{\tau}\approx 2.4\times 10^{-4},
\end{split}
\end{equation}
\begin{equation}
\label{eqn:BRm}
\begin{split}
&\mathcal{B}^D_{\leftrightarrow}(\ell=\mu)=\Gamma^{D}_{\leftrightarrow}/\Gamma_{\mu}\approx 1.9\times10^{-4} ,\\
&\mathcal{B}^M_{\leftrightarrow}(\ell=\mu)=\Gamma^{M}_{\leftrightarrow}/\Gamma_{\mu}\approx1.1\times10^{-4},
\end{split}
\end{equation}
which, in principle, is a result that could motivate even more its search and reflect the advantages of this specific process ($\ell\to\ell'\nu
\bar{\nu}
\gamma$), since we do not have to deal with hadronic form factors and the computed BR is much larger than the ones estimated in \cite{Kim} for $B$ decays ($\mathcal{B}_{\leftrightarrow}\approx 10^{-12}$) and related processes.

Nevertheless, the estimated BRs (\ref{eqn:BRt}, \ref{eqn:BRm}) seem troublesome. First of all, they are different for Dirac and Majorana cases (because we used the condition $\phi=0$, as stressed), which disagrees with the DMCT theorem, as we must integrate over the neutrinos variables to calculate them. Even so, this can be easily corrected, just considering the full range of variation of $\phi$.

The main problem is quite clear: Since the total BR($\ell\to\ell'\nu
\bar{\nu}
\gamma$) is of the order $\mathcal{O}(10^{-2}
)$, it is hard to believe that a specific kinematic configuration, such as the $b2b$, is only two orders of magnitude more suppressed than the general case. Then, for a complete discussion of this problem we will comment on the specific reason why this BR estimation is wrong, and also do the right computation for this case, which can be applied for any other specific kinematic configuration in which one (or more) of the continuous phase space variables is(are) fixed to specific value(s).

The width of an N-body decay can be seen as the hypervolume of the phase space weighted by the dynamic condition (squared amplitude) of the specific process. This hypervolume is determined by the specific range of all the continuous phase space independent variables, which is specified by the minimum and maximum values they could take according to energy-momentum conservation. If one (or more) of these variables take(s) a fixed value in a specific kinematic configuration, an integration over a zero-range variable has to be done in order to compute the theoretical BR, leading to a vanishing contribution for this specific case.

In other words, once a continuous phase space variable is fixed, the phase-space hypervolume is reduced to a phase-space hypersurface, meaning that in that case the purely theoretical BR estimation will be zero for that configuration. We note this is congruent with the intuitive notion of obtaining a null probability for a unique configuration among all the continuous (infinite) possibilities.

This does not mean that the differential decay rate is zero in that case. Actually, we can compute without further difficulties any differential distribution as long as the fixed variables are not integrated. Then, the main problem of the estimated BR is that we integrated over the already evaluated differential decay rate, leading to a number that does not have a probability interpretation, since for the correct theoretical BR computation, we first need the differential decay rate for the general case and then to integrate over all phase-space variables, which range will be fixed by the specific kinematic scenario and the energy-momentum conservation. 

In particular our notation first introduced in eqs.(\ref{neutrinospectrum}), was precisely motivated to avoid this possible confusion. It provides evidence that, once the decay rate is already evaluated in a specific kinematic scenario, we can not integrate over the kinematic variables fixed by the $b2b$ condition and interpret the result as a probability, specifically as the BR of the $b2b$ case, that could lead to a larger value than the real one.

Finally, we know that the experimental resolution is finite and thus experimentally we can not have a strictly fixed variable. Then, it is well-defined to estimate a non-zero BR for the $b2b$ configuration, considering a small range of variation for the theoretically fixed variables, according to the experimental resolution.

Then, to estimate the real BR for the experimental $b2b$ case, without fixing $\phi$ and using a proper method for this estimation, we have to do the following: First we integrate over the remaining neutrino's variables ($\theta_{\nu}$ and $\phi$) in eq.(\ref{diffwidth}), which are not fixed by any kinematic condition, leading to a decay rate that depends only on the electron and photon energy, together with the angle between them $\left(\frac{d\Gamma}{dE_{\ell'}dE{\gamma}d\cos{\Theta_{\ell'\gamma}}}\right)$, in the general kinematic case. 

\begin{figure}[tbp]
     \centering
     \includegraphics[scale=.6]{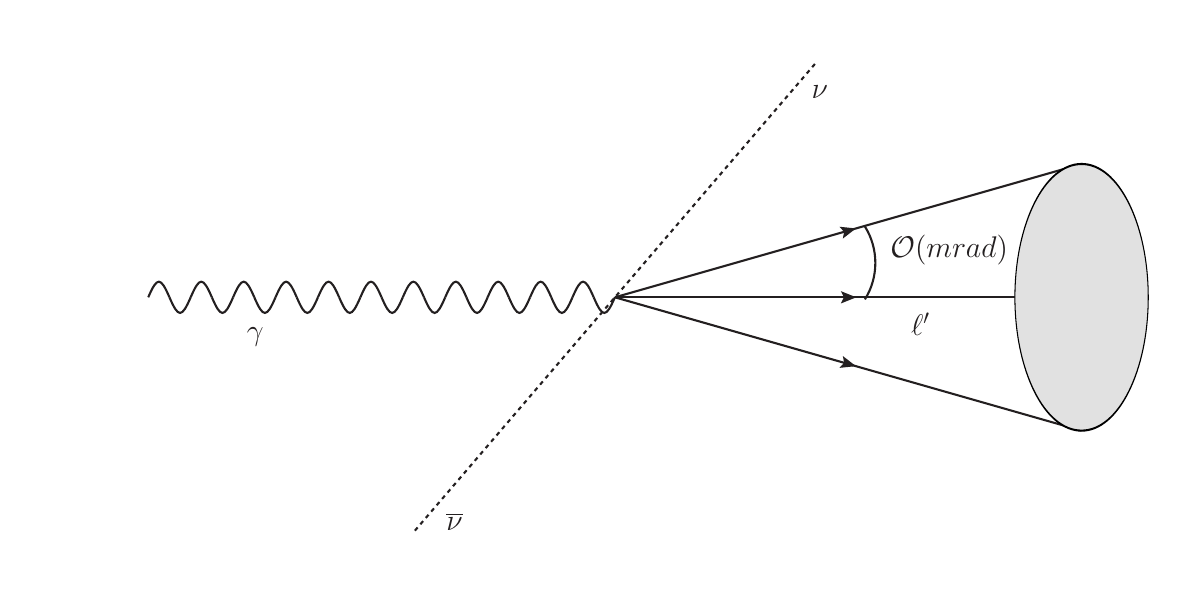}
     \caption{Experimental $b2b$ kinematic configuration in the initial charged-lepton rest frame.}
     \label{expb2b}
 \end{figure}

Then, using this differential decay rate, we can apply the experimental energy and angular resolution\footnote{We use $\Delta E$=0.01$E_{\ell'}$, $\Delta\theta$=10mrad and $\Delta E$=0.04$E_{\ell'}$, $\Delta\theta$=13mrad for the muon and tau experimental resolution respectively, as reported by Mu2e \cite{Mu2e:2014fns} and Belle-II \cite{Belle-II:2018jsg} experiments.} to integrate over the “pseudo” $b2b$ case, i.e. an infinitesimal phase-space region that will be indistinguishable from the theoretical $b2b$ case by the experiment, as shown in figure \ref{expb2b}, where the energy of the final charged-lepton and the photon are equal, up to the energy resolution ($E_{\ell'}-\Delta E\leq E_{\gamma}\leq E_{\ell'}+\Delta E$), and the angle between them is $\pi$, up to the angular resolution ($\pi-\Delta\theta\leq\theta_{\ell'\gamma}\leq\pi$).\\
Since we are not evaluating the differential decay rate in a specific kinematic configuration in any moment, we can safely interpret this result as an occurrence probability. Furthermore, since the integration region contains the theoretical $b2b$ case, the corresponding BR obtained after the integration must therefore be larger than the theoretical $b2b$ case alone.

Doing these, cutting off photons below an energy threshold of 10 MeV in the decaying-lepton rest frame,  we estimate the following BR for the experimental $b2b$ case as follows:
\begin{equation}
   \left(\Gamma\right)_{b2b}=\int_{E_{\ell'}^ {min}}^{E_{\ell'}^ {max}}\int_{E_{\ell'}-\Delta E}^{E_{\ell'}+\Delta E}\int_{\pi-\Delta\theta}^{\pi} \frac{d\Gamma}{dE_{\ell'}dE_{\gamma}d\cos{\Theta_{\ell'\gamma}}}dE_{\ell'}dE_{\gamma}d\cos{\Theta_{\ell'\gamma}},
\end{equation}
getting the results:
\begin{equation}
\label{eqn:BRbueno}
\begin{split}
&\mathcal{BR}(\ell=\tau)_{b2b}=\left(\Gamma\right)_{b2b}/\Gamma_{\tau}\approx 1.68\times 10^{-10},\\
&\mathcal{BR}(\ell=\mu)_{b2b}=\left(\Gamma\right)_{b2b}/\Gamma_{\mu}\approx 1.34\times 10^{-10}.
\end{split}
\end{equation}
As we can see, these branching ratios are many orders of magnitude smaller than the ones obtained, using the KMS method, in (\ref{eqn:BRt}) and (\ref{eqn:BRm}). Also, the BR calculated is the same for Dirac and Majorana neutrinos, in agreement with the DMCT theorem, being all consistent with our previous discussions.


\end{document}